\documentclass[12pt]{article}
\usepackage{amsfonts}
\usepackage{amsmath}
\usepackage{amssymb}
\usepackage{epsfig}
\usepackage{color}
\usepackage{float}
\usepackage{tabularx}
\usepackage{ulem}
\usepackage{cancel}
\usepackage{braket}

\definecolor{darkblue}{rgb}{0,0,.5}

\begin{document}
\thispagestyle{empty}

\begin{flushright}
November 2025\\
\end{flushright}
\vspace{5mm}
\begin{center}
\Large {\bf On the Relations between Fermion Masses and Isospin Couplings in the Microscopic Model}\\
\mbox{ }\\
\normalsize
\vspace{0.9cm}
{\bf Bodo Lampe} \\              
\vspace{0.3cm}
II. Institut f\"ur theoretische Physik der Universit\"at Hamburg \\
Luruper Chaussee 149, 22761 Hamburg, Germany \\

\vspace{1.2cm}


{\bf Abstract}
\end{center} 
\vspace{-0.5cm}
Quark and lepton masses and mixings are considered in the framework of the microscopic model. The most general ansatz for the interactions among tetrons leads to a Hamiltonian $H_T$ involving Dzyaloshinskii-Moriya (DM), Heisenberg and torsional isospin forces. Diagonalization of the Hamiltonian provides for 24 eigenvalues which are identified as the quark and lepton masses. While the masses of the third and second family arise from DM and Heisenberg type of isospin interactions, light family masses are related to torsional interactions among tetrons. Neutrino masses turn out to be special in that they are given in terms of tiny isospin non-conserving DM, Heisenberg and torsional couplings.\\ 
The approach not only leads to masses, but also allows to calculate the quark and lepton eigenstates, an issue, which is important for the determination of the CKM and PMNS mixing matrices. Compact expressions for the eigenfunctions of $H_T$ are given. The almost exact isospin conservation of the system dictates the form of the lepton states and makes them independent of all the couplings in $H_T$. Much in contrast, there is a strong dependence of the quark states on the coupling strengths, and a promising hierarchy between the quark family mixings shows up. 

\newpage

\normalsize






\begin{center}
{\bf I. Introduction}
\end{center}


Our universe according to the microscopic model\cite{bodoreview} is a 3-dimensional elastic substrate expanding within some higher dimensional space. The elastic substrate is built from tiny invisible constituents, called tetrons, with bond length about the Planck length and binding energy the Planck energy. Tetrons transform under the fundamental spinor representation of SO(6,1). This representation is complex, 8-dimensional and sometimes called the octonion representation\cite{slansky,ross}. 

Details of the approach provide a powerful unified picture for particle physics and cosmology. All physical properties in the universe can be derived from properties of the tetrons. This philosophy is applied here to the Standard Model mass and mixing parameters which are shown to be determined by the interactions among tetrons. 

The 24 known quarks and leptons arise as eigenmode excitations of a tetrahedral fiber structure, which is made up from 4 tetrons and extends into 3 extra `internal' dimensions. While the laws of gravity are due to the elastic properties of the tetron bonds\cite{bodogravity}, particle physics interactions take place within the internal fibers, with the characteristic internal energy being the Fermi scale. All ordinary matter quarks and leptons are constructed as quasiparticle excitations of this internal fiber structure. Since the quasiparticles fulfill Lorentz covariant wave equations, they perceive the universe as a 3+1 dimensional spacetime continuum. 

More in detail, the ground state of our universe looks like illustrated in Fig. 1. In this figure the tetrahedrons (=`fibers') extend into the 3 extra dimensions. The picture is a little misleading because in the tetron model physical space and the extra (`internal')  dimensions are assumed to be completely orthogonal. This means the whole game has to be played within a larger, at least 6 dimensional space, 3 physical dimensions and 3 internal ones\footnote{There are indications that the system actually lives in 7+1 dimensions instead of 6+1; this does not alter the calculations and results presented in this article, but may be important for an overall understanding of the structure of the universe. Details are discussed in Appendix D.}.

Each tetrahedron in Fig. 1 is made up from 4 tetrons, depicted as dots. With respect to the decomposition of $SO(6,1)\rightarrow SO(3,1)\times SO(3)$ into the (3+1)-dimensional base space and the 3-dimensional internal space, a tetron $\Psi$ possesses spin $\frac{1}{2}$ and isospin $\frac{1}{2}$. This means it can rotate both in physical space and in the extra dimensions, and corresponds to the fact that $\Psi$ decomposes into an isospin doublet $\Psi=(U,D)$ of two ordinary SO(3,1) Dirac fields U and D.
\begin{eqnarray}
SO(6,1)&\rightarrow& SO(3,1)\times SO(3) \nonumber \\
8 &\rightarrow& (1,2,2)+(2,1,2)=((1,2)+(2,1),2)
\label{eq8}
\end{eqnarray}

\begin{figure}[p]
\begin{center}
\includegraphics[width=5.0in]{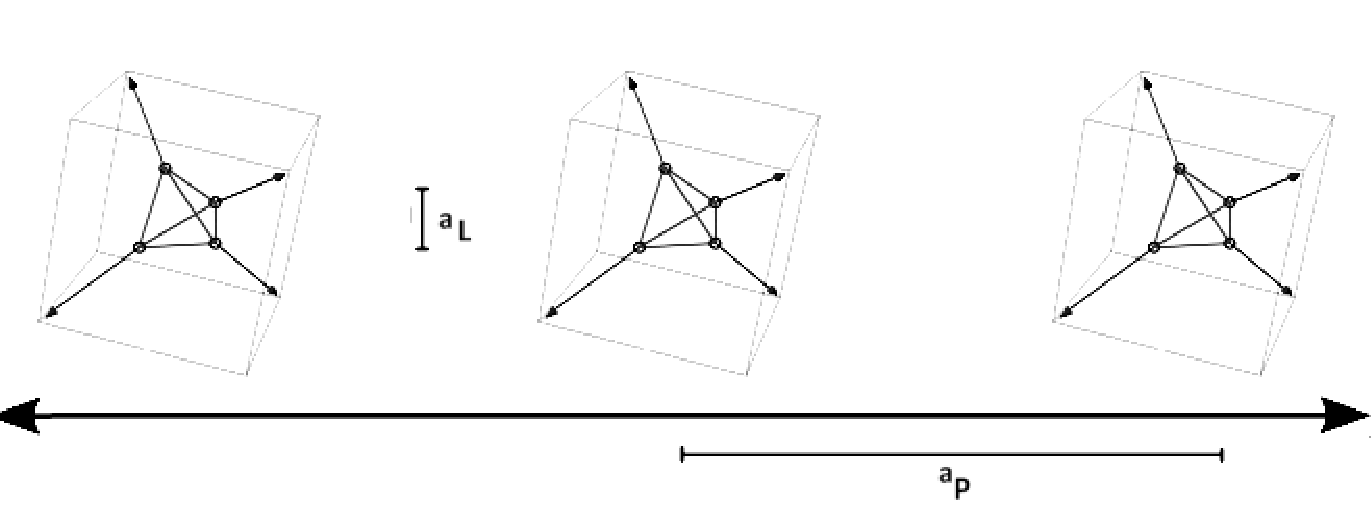}
\end{center}
\caption{The global ground state of the universe after the electroweak symmetry breaking has occurred, considered at Planck scale distances. It consists of an aligned system of tetrahedrons each extending into 3 extra dimensions. The big black double arrow represents 3-dimensional physical space. $a_L$ is the magnitude of one tetrahedron within the 3 extra dimensions and $a_P$ the average distance between two neighboring tetrahedrons. The small arrows are the isospin vectors $\vec Q_L+\vec Q_R$ defined in (\ref{eq894}) and used in (\ref{mm3}). Before the symmetry breaking the isospin vectors are directed randomly, thus exhibiting a local SU(2) symmetry, but once the temperature drops below the Fermi scale $v_F$, they become ordered into a repetitive tetrahedral structure, thereby spontaneously breaking the initial SU(2). Note that the SM Higgs vev $v_F$ is related to the length of the aligned isospin vectors in neighboring tetrahedrons. The figure is therefore a bit misleading, not only because the tetrahedrons do not have an extension into physical space, but also the relative magnitudes are not correctly drawn. 
While $a_L$ and $a_R$ are of the order of the Planck length, the extension of the tetrahedrons formed by the isospin vectors is dictated by the Fermi scale. 
While gravity can be attributed to the elasticity of the coordinate bonds\cite{bodogravity}, the phenomena of particle physics arise from the interactions between isospin vectors. 
The figure shows how our universe looks like in the tetron model. It is part of a 6-dimensional space and is a 3-dimensional 'monolayer' of tetrahedrons each extending into the remaining 3 extra dimensions. The monolayer ground state acts as a background on which quarks and leptons glide as quasiparticle excitations. It has the properties of a Lorentz ether and is thereby not in conflict with Michelson-Morley type of experiments.}
\label{aba:neufig1}
\end{figure}

Why this tetrahedral structure? It is needed in order to explain the observed quark and lepton spectrum, which means to get exactly 24 excitation states with the correct multiplet structure\footnote{The quark triplets are triplets under tetrahedral transformations at this point. For the question how to interpret them as SU(3) color triplets one may consult Appendix B.}. In fact, the tetrahedral symmetry is rather uniquely determined by this condition\cite{bodohiggs}. As shown below, under reasonable assumptions on the tetron dynamics, the numerical mass values of quarks and leptons can be correctly reproduced.

The arrows in Fig. 1 denote the isospins, i.e. internal spin vectors of the tetrons. More precisely, each arrow stands for two(!) vectors $\langle \vec Q_L\rangle = \langle \vec Q_R\rangle$ where\cite{pcacpaper}
\begin{eqnarray}
\vec Q_L=\frac{1}{4}\Psi^\dagger (1-\gamma_5)\vec \tau\Psi =\frac{1}{2}\Psi^\dagger_L \vec\tau \Psi_L
\qquad \quad
\vec Q_R= \frac{1}{4}\Psi^\dagger (1+\gamma_5)\vec \tau\Psi=\frac{1}{2}\Psi^\dagger_R \vec\tau \Psi_R 
\label{eq894}
\end{eqnarray}
and $\langle \rangle$ denotes the ground state/vacuum expectation values (i.e. after the SSB). In other words, the ground state values $\langle \vec Q_{Li}\rangle$ and $\langle \vec Q_{Ri}\rangle$ are assumed to be equal on each tetrahedral site i=1,2,3,4 and given by one of the arrows in Fig. 1. $\vec \tau$ are the internal spin Pauli matrices.  

According to (\ref{eq8}) the tetron representation 8 contains both particle and antiparticle degrees of freedom. $\vec Q_L$ and $\vec Q_R$ cover 6 of its 8 dof\footnote{The remaining 2 dof correspond to the `densities' 
\begin{eqnarray}
n_L=\frac{1}{4}\Psi^\dagger (1-\gamma_5)\Psi =\frac{1}{2}\Psi^\dagger_L  \Psi_L
\qquad \quad
n_R= \frac{1}{4}\Psi^\dagger (1+\gamma_5)\Psi=\frac{1}{2}\Psi^\dagger_R \Psi_R 
\label{pm11gg92}
\end{eqnarray}
whose fluctuations actually are dark matter candidates\cite{bodoreview}.}. Furthermore, $\vec Q_L$ and $\vec Q_R$ are particularly useful to handle because quantum mechanically they commute with each other\cite{pcacpaper}. As turns out, the interactions of these internal spins play an essential role for particle physics and for electroweak symmetry breaking. 

Due to the pseudovector property of the isospin vectors their tetrahedral symmetry group actually is a Shubnikov point group\cite{white}. This means, while the coordinate symmetry is $S_4$, the arrangement of isospin vectors respects the tetrahedral Shubnikov symmetry\cite{crack}
\begin{eqnarray}
G_4:=A_4+CPT(S_4-A_4)
\label{eq8gs}
\end{eqnarray}
where $A_4 (S_4)$ is the (full) tetrahedral symmetry group and CPT the usual CPT operation except that P is the parity transformation in physical space only. Since the elements of $S_4-A_4$ contain an implicit factor of internal parity, the symmetry (\ref{eq8gs}) certifies CPT invariance of the local ground state in the full of $R^{6+1}$. Nevertheless, by itself each tetrahedron of isospin vectors is chiral within the 3 extra dimensions, the configuration with opposite internal chirality being given when the isospin vectors would point inwards instead of outwards. 
Note that in the situation depicted in Fig. 1 the SU(2) symmetry breaking has already occurred, because the isospins are aligned between all the tetrahedrons. Before the symmetry breaking, which means above a certain temperature, isospins are directed randomly, corresponding to a local $SU(2)\times U_1$ symmetry\footnote{Weak parity violation, vulgo the appearance of index L in $SU(2)_L$, arises from the chirality of the isospin tetrahedrons, because their internal handedness is dynamically related to the $V-A$ nature of the weak interaction. This, as well as the Z-$\gamma$ mixing, is discussed in detail in \cite{bodoreview}.}, but when the universe cools down, there is a phase transition, and the isospins freeze into the aligned structure, breaking the $SU(2)$ symmetry to the discrete `family group' $G_4$. And the important point to note is, this temperature can be identified with the Fermi scale\cite{bodohiggs}. Moreover, the remaining symmetry $G_4 \times U(1)_{em}$ is valid down to the lowest energies.

As elaborated in the following sections, the mathematical treatment of the excitations arising from the isospin interactions (\ref{txxm32}) and (\ref{mm3444}) is similar to that of magnons in ordinary magnetism. However, the physics is quite different, because in contrast to magnons the isospin excitations are pointlike, i.e. they can exist within one point of physical space, because they are vibrations of the isospin vectors of the tetrons within one internal tetrahedron. Note, that these internal vibrations  are spin-$\frac{1}{2}$ because they inherit their fermion nature from the fermion property of the vibrating tetrons in their 3-dimensional physical `base space'.  

Similar to magnons, the vibrations can move in physical space\cite{white} by hopping from one tetrahedron to another (particle picture) or propagating as quasiparticle waves through physical space (wave picture). Thus, although it can exist at one point of physical space, when one tries to exactly measure its location, for example by scattering with another particle, the excitation will start to move on physical space, and this movement will follow a wave equation which naturally has an uncertainty in it according to Schwarz' inequality. Planck's constant enters this uncertainty because the whole process is taking place on a discrete system with Planck length 'lattice constant' and Planck energy 'response energy'. 

The second part of the article deals with the mixing of families and the question to what extent it can be deduced from tetron ideas. Since as much as 8 of the 19 free SM parameters arise from those mixings, there have been many attempts to reduce this freedom by BSM ideas. That is the reason why in the literature a lot of suggestions for relations among the CKM resp PMNS matrix elements can be found, e.g. \cite{mix1,mix2,mix3,mix4}, mostly on the basis of assumptions on additional discrete symmetries, from which such relations then are derived. 

\begin{center}
{\bf II. Quark and Lepton Masses from the Interactions of Isospins}
\end{center}

The SM SSB being realized by an alignment of the tetron isospins, it is not surprising that the masses of quarks and leptons, and thus the SM Yukawa couplings are determined by the interactions among those isospins. The simplest interaction Hamiltonian between isospin vectors of 2 tetrons i and j looks like 
\begin{equation} 
H=- J \,  \vec Q_{i}\vec Q_{j}
\label{mm3}
\end{equation}
So it has the form of a Heisenberg interaction - but for isospins, not for spins. The coupling J may be called an `isomagnetic exchange coupling'. Note that the language of magnetism often is used in this paper, although interactions of isospins and not of spins are considered. Note further that isospin is not an abstract symmetry here, but corresponds to real rotations in the 3 extra dimensions.

In reality, the Hamiltonian H is more complicated than (\ref{mm3}), for several reasons:\\
$\bullet$ There are inner- and inter-tetrahedral interactions of isospins, i.e. within the same and with a neighboring tetrahedron. The inner ones must have an energy minimum at the tetrahedral angle $\theta=\theta_{tet}=\arccos(-\frac{1}{3})$, while the inter ones correspond to a minimum at the collinear configuration $\theta=0$, cf. Fig. 1.\\
$\bullet$ The appearance of antitetron degrees of freedom should be accounted for by using interactions both of $\vec Q_L$ and $\vec Q_R$ defined in (\ref{eq894}) instead of $\vec Q$ in (\ref{mm3}). The Heisenberg Hamiltonian for the interaction between 2 tetrons i and j then reads:
\begin{equation} 
H_H=- J_{LL}\,  \vec Q_{Li}\vec Q_{Lj}- J_{LR}\,  \vec Q_{Li}\vec Q_{Rj}- J_{RR}\,  \vec Q_{Ri}\vec Q_{Rj}
\label{mmHL3}
\end{equation}
As shown later in Sect. IV, the three couplings $J_{LL}$, $J_{LR}$ and $J_{RR}$ can be roughly associated to the masses of the second family fermions, $m_c$, $m_\mu$ and $m_s$, respectively.\\ 
$\bullet$ In addition to the Heisenberg Hamiltonian (\ref{mmHL3}) Dzyaloshinskii interactions\cite{dm} are to be considered. They will be shown to give the dominant mass contributions to the heavy family. As well known, the form of the DM couplings $\vec D_{ab}$ in (\ref{mm3444}) is restricted by the ground state symmetry through the so-called Moriya rules\cite{moriya}. Applying these rules to the given tetrahedral structure, the DM Hamiltonian can be shown to have the form (\ref{mmDM3}).\\
$\bullet$ Heisenberg and DM terms do not contribute at all to the masses $m_e$, $m_u$ and $m_d$ of the first family. Therefore, small torsional interactions are introduced in Sect. V. They are characterized by the exerting torques $dQ_{L,R}/dt$ being proportional to the isospins $Q_{L,R}$ themselves, cf. Eq. (\ref{hoo2}).\\ 
$\bullet$ The masses of the neutrinos are yet another story. While the interactions discussed so far are isospin conserving and leave the neutrinos massless, neutrino masses can arise only from isospin violation. Generation of these masses will be discussed in Sect. VI, and a physical explanation for the origin of the isospin violation will be given.

The DM-couplings $K_{LL}$, $K_{LR}$ and $K_{RR}$ introduced in (\ref{mmDM3}) are much larger than both Heisenberg and torsional interactions and essentially determine the masses $m_\tau$, $m_b$ and $m_t$ of the third family particles. $K_{LL}$ will be shown to be particularly large. It gives the dominant contribution to the top mass and is the main source for the parallel arrangement of isospins and the SU(2) SSB.

All the types of interaction mentioned above contribute to the angular dependence of the 
energy of 2 tetron isospins at angle $\theta$ which is basically of the form
\begin{equation} 
E=A +B \,\cos(\theta) +C \,\cos^2(\theta)
\label{mmXX3}
\end{equation}
where $A$, $B$ and $C$ are determined by the Heisenberg, torsional and DM coupling strengths. (For example, the Heisenberg coupling $J_{LL}$ in (\ref{mmHL3}) concerns $\theta_{LL} = \sphericalangle (\vec Q_{Li},\vec Q_{Lj})$ and gives a contribution to $B_{LL}$ only.) Altogether, they fix the {\it relative} directions of the ground state isospins at the energy minimum, both locally and globally in the way depicted in Fig. 1 - whereas the {\it absolute} arrangement of the tetrahedrons is spontaneous.

Furthermore,  they give rise to fermionic excitations which are to be interpreted as quarks and leptons. Masses can then be calculated using the Hamiltonians discussed above. Indeed, 24 eigen energies arise from the tetrahedral configuration by diagonalizing equations for the isospin torque which are generically of the form
\begin{equation} 
\frac{d\vec Q}{dt} = i \, [H, \vec Q] 
\label{txxm32}
\end{equation}
While the masses correspond to the eigenvalues, CKM and PMNS mixings can be deduced from the eigenvectors. This point will be discussed in Sects. VII ff.

More in detail, the quarks and leptons are vibrations $\delta$ of the isospin vectors $\vec Q_{Li}$ and $ \vec Q_{Ri}$ of the tetrons $i$ at sites $i=1,2,3,4$, i.e. fluctuations of the ground state values within one tetrahedron. 
\begin{eqnarray}
\vec Q_{Li}=\langle \vec Q_{Li}\rangle+\, \vec \delta_{Li} \qquad \qquad \qquad \vec Q_{Ri}=\langle \vec Q_{Ri}\rangle+\, \vec \delta_{Ri}
\label{eqxxdrt1}
\end{eqnarray}
where 
\begin{eqnarray}
\langle \vec Q_{Li} \rangle
=\frac{1}{4}\langle \Psi^\dagger (1-\gamma_5)\vec \tau \Psi\rangle  
\qquad \qquad
\langle \vec Q_{Ri}\rangle = \frac{1}{4}\langle \Psi^\dagger  (1+\gamma_5)\vec \tau \Psi\rangle 
\label{eqtzt1}
\end{eqnarray}
are the ground state radial isospin vectors of a tetrahedron in Fig. 1 assumed to be pointing outward
\begin{eqnarray}
\langle \vec Q_{Li} \rangle = \langle \vec Q_{Ri}\rangle = \vec e_r
\label{eqtzt1h}
\end{eqnarray}

In the present model, gauge and Higgs bosons are constructed as excitations of a single pair of a tetron and an antitetron belonging to 2 neighboring tetradedrons\footnote{Note that these excitations are bosons because they involve tensor products of 2 tetron fields over 2 points in physical space. This point is discussed in more detail in Appendix C.}. As excitations of single pairs they do not feel their tetrahedral environment, which means they transform under isospin but do not appear in multiplets of the tetrahedral group.

One may rightfully ask whether the Higgs mass and in general the parameters of the SM Higgs potential can be determined from similar considerations as the fermion masses. I do not have a complete answer to this question, but will draw below an analogy with a ferromagnet. Actually, this is a very meaningful analogy because in the microscopic model the `isomagnetic' alignment between isospin vectors of neighboring tetrahedrons plays an important role for the SSB of SU(2). 

Within the SM the mass of the Higgs particle is given by
\begin{eqnarray}
m_H=\sqrt{2}|\mu|    
\label{eqx4t1}
\end{eqnarray}
where $\mu$ is defined through the potential of the Higgs doublet $\phi$
\begin{eqnarray}
V(\phi)=-\mu^2 \phi^\dagger \phi + \lambda (\phi^\dagger \phi)^2
\label{eqx4t12}
\end{eqnarray}
It is instructive to compare this to the free energy of a ferromagnet given in terms of its magnetization M
\begin{eqnarray}
F=U-TS=\alpha \vec M^2+ \beta \vec M^4  
\label{eqx4t22}
\end{eqnarray}
Here $\alpha$ gets a negative contribution from the Heisenberg interaction energy U
between 2 spins and a positive contribution from the temperature/entropy term. 
The transition energy is then obtained from the condition $\alpha=0$ as $kT_C=2U$, i.e. the transition energy is directly related to the interaction between 2 spins. 

This may be compared to the present model: for the tetrahedrons to align as in Figure 1, pairs of isospin vectors from neighboring tetrahedrons must align. As described after (36), the second term in (36) is a Heisenberg type of `ferromagnetic' coupling with strength $\frac{2}{3} K_{LL}+J_{LL}$ (where $K_{LL}\gg J_{LL}$). This means the coupling $K_{LL}$ is the dominant driving source for the SSB alignment, and at the same time according to (34) the coupling $K_{LL}$ gives the dominant contribution to the top mass. One is thus led to the conclusion that $m_H$ and $m_t$ are roughly of the same order of magnitude and, more specifically, that the Higgs mass should be about two thirds of the top mass.

More details about how the SM Higgs and gauge fields are to be interpreted within the microscopic model can be found in Appendix C.


\begin{center}
{\bf III. Physical Origin of the Isospin Interactions}
\end{center}

This chapter is devoted to the question how the Heisenberg, DM and torsional interactions introduced in the last section can be understood from a more fundamental interaction among tetrons.

First of all, the interested reader should remember that Heisenberg used the Heitler-London results for the hydrogen molecule to understand the phenomenon of ferromagnetism. Heisenberg showed\cite{heisenberg} that ferromagnetism is a quantum effect arising from the Pauli principle, more precisely, from the large exchange energies due to the overlap of the antisymmetrized electron wave functions.

The situation here is in principle similar - but in practice somewhat more complicated, because one deals with 6 dimensions with 2 types of rotations: spin and isospin. 

In the non-relativistic limit $SO(6,1)\rightarrow SO(6)$ the tetron representation $8$ of SO(6,1) reduces to 
\begin{eqnarray}
SO(6,1)&\rightarrow& SO(6) \\
8&\rightarrow& 4+\bar{4}
\label{eqrt1vv}
\end{eqnarray}
where $4$ is the spinor representation of $SO(6)$ and $\bar 4$ its complex conjugate. Since the universal covering of SO(6) is given by SU(4), the $4$-representation actually is the fundamental representation of SU(4). This representation contains the spin ($\pm \frac{1}{2}$) and isospin ($\pm \frac{1}{2}$) of the tetron, while the $\bar 4$-representation corresponds to the antitetron degrees of freedom. 

Within a non-relativistic quantum mechanics the binding energy between a tetron and an (anti)tetron should generally be calculable from the expectation value 
\begin{eqnarray}
E_F=\int d^6 x_i d^6 x_j \Phi_F^* (x_i,x_j) U_F(|x_i - x_j|) \Phi_F (x_i,x_j)  
\label{eqrt2}
\end{eqnarray}
of a non-relativistic potential $U_F$, where $\Phi$ is the complete wavefunction for the tetron-(anti)tetron system and $F$ denotes its combined list of quantum numbers, i.e. spin, isospin, orbital angular momentum etc. 

$\Phi_F$ may be approximated by a sum of products of two 1-tetron wave functions concentrated at the two tetrahedral sites $x_i$ and $x_j$. 
Antisymmetrization of this sum will lead $E_F$ to consist of two terms, the classical `direct' integral $D_F$ and the quantum mechanical exchange contribution $J_F$. 
\begin{eqnarray}
E_F=D_F + J_F  
\label{eqrt2h}
\end{eqnarray}
While $D_F$ determines the elastic binding among tetrons and thus the gravitational properties of the substrate, the exchange integral $J_F$ can be used to understand the isospin interactions and thus the phenomena of particle physics. Actually, as seen below, $J_F$ is directly related to the `isomagnetic' Heisenberg and DM couplings J and K defined in the last section.

If one assumes the single tetron wave functions to be fairly localized at their tetrahedral sites, there is a hierarchy $|J_F|\ll |D_F|$. This is different from ordinary 3-dimensional ferromagnetism and is even enhanced by the 12-dimensional integration in (\ref{eqrt2}), through which any overlap contribution becomes strongly suppressed as compared to a direct one. In the extreme case of delta functions, $D_F$ reflects the form of the potential, while $J_F$ vanishes. In the general case, $D_F$ will still be much larger than $J_F$. For example, assuming the single tetron wave function to fall off by a factor of 10 at half the distance between the 2 sites i and j, $J_F$ will be smaller than $D_F$ roughly by a factor of $10^{-12}$. This, en passant, is the way the hierarchy between the Planck scale and the Fermi scale can be understood within the tetron approach. The item has been discussed more thoroughly in \cite{bodogravity}. 

One may ask how the potential $U_F$ transforms under SU(4). Since the energy of 2 tetrons must be a singlet, one has to have 
\begin{eqnarray}
(4+\bar 4)\times R_U \times (4+\bar 4) = 1+...
\label{eqrt3}
\end{eqnarray}
where $R_U$ is the representation under which $U_F$ transforms. Since $4\times \bar 4=1+15$ and $4\times 4=6+10$ and $15\times 15=1+...$ and $6\times 6=1+...$\cite{slansky}, it follows that $U_F$ is either a scalar $U_1$,  an adjoint $U_{15}^a \lambda^a$, a=1,...,15, or a vector $U_6^i e^i$, i=1,...,6, where $\lambda^a$ are the generators of SU(4) and $e^i$ are vectors which span 6-dimensional space. $U_1$ and $U_{15}$ describe interactions among a tetron and an antitetron and $U_6$ is a tetron-tetron interaction. 

In the present context, where the tetrahedrons are completely orthogonal to physical space, spin and isospin essentially decouple from each other, and the above analysis may be strongly simplified, in the following way: instead of (\ref{eqrt1vv}) one may consider
\begin{eqnarray}
SO(6,1)&\rightarrow& SO(3)_{spin}\times SO(3)_{isospin} \nonumber \\
8&\rightarrow& (1,2)+(1,2)+(2,1)+(2,1)
\label{eqrt198}
\end{eqnarray}
Since the ordinary spin of the tetrons (i.e. the spin in physical space) is irrelevant for the internal interactions, it is enough to look for SO(3)$_{isospin}$ singlets in $2\times R\times 2=1+...$, which implies $R=1$ or $R=3$, i.e. only an isospin singlet or a triplet potential $V_1$ or $V_3$ are allowed for the isospin interactions among tetrons.



$V_1$ and $V_3$ may be considered as part of the above $SO(6)$ potentials $U_1$ and $U_{15}$ and induced by them within the $SO(3)_{isospin}$ fibers. Alternatively, $V_1$ and $V_3$ can also be shown to arise in the relativistic framework, i.e. sticking to the original octonion representation 8 of SO(6,1) instead of using (\ref{eqrt1vv}). Namely, a relativistic potential $W_7$ is allowed that transforms as 7 under SO(6,1), and the product\cite{slansky}
\begin{eqnarray}
8\times 7\times 8=1+7+7+21+21+27+35+35+105+189
\label{eq19}
\end{eqnarray}
contains a singlet. 

$W_7$ may well be a gauge potential and the basis for the fundamental tetron interaction. Furthermore, $V_1$ and $V_3$ are part of $W_7$ due to 
\begin{eqnarray}
SO(6,1)&\rightarrow& SO(3)_{spin}\times SO(3)_{isospin} \\
7 &\rightarrow& (1,1)+(1,3)+(3,1)
\label{eqrt15a}
\end{eqnarray}
where $V_1$ transforms as (1,1) and 
\begin{eqnarray}
V_3=V_{3}^a \tau^a=\vec V_3 \, \vec \tau
\label{eqrt15d}
\end{eqnarray}
as the isospin triplet (1,3)\footnote{In contrast to the suggestion in \cite{bodohiggs} the SM photon should {\it not} be assumed to be part of the gauge field $W_7$. As explained before, the photon as well as all the other SM gauge bosons are excitations of tetron-antitetron bonds of neighboring tetrahedrons. Nevertheless, they transform under SO(3)$\times$SO(3), the weak bosons, for example, as (3,3), i.e. they are spin 1 and isospin 1 particles.}. 

While the Heisenberg interaction $\sim Q^a_i Q^a_j$ is associated to the singlet potential $V_1$ in the usual way\cite{heisenberg}, DM terms $\sim \epsilon^{abc} Q^b_i Q^c_j$ in (\ref{mm3444}) arise from the $V_3$ contributions. This can be shown by inserting the completeness relation  for Pauli matrices
\begin{eqnarray}
\delta_{sv}\delta_{ut} = 2 \tau^a_{st} \tau^a_{uv} +\frac{1}{2}\delta_{st}\delta_{uv} 
\label{eqrt15h}
\end{eqnarray}
into the $V_3$-exchange integral and afterwards noting that the factor of $\tau^a$ in (\ref{eqrt15d}) can be merged with one of the factors $\tau$ in (\ref{eqrt15h}) via 
\begin{eqnarray}
\tau^a \tau^b = i\epsilon^{abc}\tau^c + \delta^{ab}
\label{eqrt15e}
\end{eqnarray}
The $\epsilon$ tensor part in (\ref{eqrt15e}) then directly yields the 'antisymmetric exchange'(=DM) contribution (\ref{mm3444}). 

More precisely, exchange integrals with $\vec V_3$ instead of $V_1$ correspond to the DM couplings $K \vec D$ in (\ref{mm3444}).


\begin{center}
{\bf IV. Dzyaloshinskii Masses for the Heavy Family; Heisenberg Masses for the Second Family}
\end{center}

My presentation of the mass calculations begins with the Dzyaloshinskii-Moriya (DM) coupling, 
firstly because it is the dominant isospin interaction and secondly it gives masses only to 
the third family, i.e. to top, bottom and $\tau$, while leaving all other quarks and leptons massless. 

Among all the fermion masses the top quark mass is by far the largest and is of the order of the Fermi scale. As turns out, this is no accident, but has to do with the largeness of the relevant DM coupling.

In the simplest version the isospin DM interaction\cite{bodoreview,dm} is
\begin{equation} 
H_{DM}=- K \,  \sum_{i\neq j=1}^4 \, \vec D_{ij} (\vec Q_{i} \times \vec Q_{j} ) 
\label{mm3444}
\end{equation}
to be compared to the Heisenberg interaction (\ref{mm3}). The form of the vectors $\vec D_{ij}$ is dictated by the tetrahedral symmetry to be\cite{moriya} 
\begin{equation} 
\vec D_{ij}=\vec Q_{i}\times \vec Q_{j}
\label{mm3444a}
\end{equation}
As explained before, interactions among $\vec Q_{L}$ and $\vec Q_{R}$ have to be considered in order to cover all degrees of freedom. The complete DM Hamiltonian then reads
\begin{eqnarray} 
H_D&=&- K_{LL} \,  \sum_{i\neq j=1}^4 \,  (\vec Q_{Li} \times \vec Q_{Lj} ) ^2
- K_{LR} \,  \sum_{i\neq j=1}^4 \,  (\vec Q_{Li} \times \vec Q_{Rj} ) ^2 \nonumber \\
& &- K_{RR} \,  \sum_{i\neq j=1}^4 \,  (\vec Q_{Ri} \times \vec Q_{Rj} ) ^2 
\label{mmDM3}
\end{eqnarray}
with DM couplings (= $V_3$ exchange integrals) $K_{LL}$, $K_{LR}$ and $K_{RR}$.

It is convenient to already include at this point the Heisenberg terms
\begin{eqnarray} 
H_H&=&-J_{LL} \,  \sum_{i\neq j=1}^4 \,  \vec Q_{Li} \vec Q_{Lj} 
-J_{LR} \,  \sum_{i\neq j=1}^4 \,  \vec Q_{Li} \vec Q_{Rj} 
-J_{RR} \,  \sum_{i\neq j=1}^4 \,  \vec Q_{Ri} \vec Q_{Rj} 
\label{all34}
\end{eqnarray}
with $V_1$ exchange couplings $J_{LL}$, $J_{LR}$ and $J_{RR}$. They are smaller than the DM interactions and turn out to give masses both to the second and third family (but not to the first one). 

Phenomenologically, the Heisenberg couplings J are typically smaller than 1 GeV, while the DM couplings K are larger than 1 GeV. Altogether, Heisenberg and DM terms provide the most general isotropic and isospin conserving interactions within the internal space. Apart from that there will only be tiny torsional interactions responsible for the mass of the first family and still smaller isospin violating interactions giving masses to the neutrinos (to be discussed in Sects. V and VI).

The masses $m$ of the corresponding excitations $\delta$ defined in (\ref{eqxxdrt1}) 
arise from the exponents in the vibrations
\begin{eqnarray} 
\delta \sim\exp (imt)=\exp (i X t)
\label{all34xys}
\end{eqnarray}
where X stands for the appropriate linear combination of the isospin couplings J and K introduced in (\ref{mmDM3}) and (\ref{all34}). The combinations X will be obtained from the torque  equations (\ref{txxm32}), using the angular momentum commutation relations for the isospin vectors\cite{pcacpaper}
\begin{equation} 
[Q_{Ri}^{a},Q_{Rj}^{b}]=i\delta_{ij}\epsilon^{abc} Q_{Ri}^{c} 
\qquad
[Q_{Li}^{a},Q_{Lj}^{b}]=i\delta_{ij}\epsilon^{abc} Q_{Li}^{c} 
\qquad
[Q_{Ri}^{a},Q_{Lj}^{b}]=0 
\label{mm3444b}
\end{equation}
where $i,j=1,2,3,4$ count the 4 tetrahedral edges and $a,b,c=1,2,3$ the 3 internal  directions(=extra dimensions).

It may be stressed that I have noch undertaken to calculate the couplings J and K in terms of the 12-dimensional $V_1$ and $V_3$ exchange integrals as defined in (\ref{eqrt2h}) and  (\ref{eqrt2}). What is done here, is to use the J and K as free parameters and calculate the masses of the excitations in terms of these couplings. This is the usual approach in magnetic theories, where it often turns out that calculation of integrals like (\ref{eqrt2}) are plagued with large and uncertain corrections. Keeping the couplings as free parameters usually is more rewarding for physical applications.

When carrying out the calculation, care must be taken concerning the unique choice of a quantization axis $\vec Q_0$ \cite{elhajal}, because this is the condition under which (\ref{mm3444b}) holds. One may choose one of the tetrahedral edges, e.g. 
\begin{equation} 
\vec Q_0:\equiv \langle \vec Q_1\rangle=\frac{1}{\sqrt{3}} (-1,-1,-1)  
\label{mm3444d}
\end{equation}
to define the axis of quantization and then has to rotate the other isospins to this system. 


The 24 first order differential equations for $dQ/dt$ arising from $H_H$ and $H_D$ are rather lengthy. In linear approximation they read 
\begin{eqnarray} 
\frac{d\vec \delta_{Li}}{dt}&=&  2 K_{LL}  \{ \vec Q_{0}\times \vec \Delta_{LLi} +i [-\vec \Delta_{LLi} +(\vec \Delta_{LLi} .\vec Q_0) \, \vec Q_0]\} \nonumber \\
&+& 2 K_{LR}  \{ \vec Q_{0}\times \vec \Delta_{LRi} +i [-\vec \Delta_{LRi} +(\vec \Delta_{LRi} .\vec Q_0) \,\vec Q_0]\} \nonumber\\
&+& J_{LL}  ( \vec Q_{0}\times \vec \Delta_{LLi}) + J_{LR}  ( \vec Q_{0}\times \vec \Delta_{LLi}) 
\label{allnxxg} \\
\frac{d\vec \delta_{Ri}}{dt}&=&  2 K_{RR}  \{ \vec Q_{0}\times \vec \Delta_{RRi} +i [-\vec \Delta_{RRi} + (\vec \Delta_{RRi} .\vec Q_0) \,\vec Q_0]\} \nonumber \\
&+& 2 K_{LR}  \{ \vec Q_{0}\times \vec \Delta_{RLi} +i [-\vec \Delta_{RLi} + (\vec \Delta_{RLi} .\vec Q_0)\, \vec Q_0]\} \nonumber \\
&+& J_{RR}  ( \vec Q_{0}\times \vec \Delta_{RRi}) + J_{LR} (  \vec Q_{0}\times \vec \Delta_{RLi} )
\label{allnxxg1}
\end{eqnarray}
In these equations $\vec \delta_{Li}=\vec Q_{Li}- \langle \vec Q_{Li}\rangle $ and $\vec \delta_{Ri}=\vec Q_{Ri}- \langle \vec Q_{Ri}\rangle $,  $a=1,2,3,4$, denote the isospin vibrations and the $\Delta$'s are certain linear combinations of them which will play an important role in discussing isospin conservation in Sect. VI:
\begin{eqnarray} 
\vec \Delta_{LLi}&=&- 3 \,\vec \delta_{Li} + \sum_{j\neq i} \vec \delta_{Lj}  \nonumber \\
\vec \Delta_{LRi}&=&- 3 \,\vec \delta_{Li} + \sum_{j\neq i} \vec \delta_{Rj}  \nonumber \\
\vec \Delta_{RLi}&=&- 3 \,\vec \delta_{Ri} + \sum_{j\neq i} \vec \delta_{Lj}  \nonumber \\
\vec \Delta_{RRi}&=&- 3 \,\vec \delta_{Ri} + \sum_{j\neq i} \vec \delta_{Ri}  
\label{anx5}
\end{eqnarray}

Eqs. (\ref{allnxxg}) and (\ref{allnxxg1}) are the basis of the Mathematica program included in Appendix A and correspond to a 24$\times$24 eigenvalue problem which - after the SSB - leads to 6 singlet and 6 triplet states of the Shubnikov group (\ref{eq8gs}), the latter ones each consisting of 3 degenerate eigenstates (corresponding to three colors, cf. Appendix B).

After diagonalization one obtains the following results: the first family excitations are still massless at this point, but will get masses from the torsional interactions to be discussed in the next section. The DM exchange coupling $K_{LL}$ is consistently of the order of the transition energy $v_F$ and the DM and Heisenberg couplings  can be accommodated to reproduce the third and second family masses. 

Namely, assuming the DM couplings K to dominate over the Heisenberg couplings J,  one can prove the following approximate relations  
\begin{eqnarray} 
m_t &=& 4 K_{LL} +O(J)\, \;\;\;\;    m_\tau=\frac{3}{2} K_{LR} +O(J)\,\;\; \;\;    m_b=4 K_{RR}  +O(J) \nonumber \\
m_c&=& J_{LL}\,\;\; \;\; \;\; \;\;    m_\mu=\frac{3}{2} J_{LR}\,\;\; \;\;\;\; \;\;     m_s=J_{RR}
\label{all36}
\end{eqnarray}
In this approximation, the masses of quarks and leptons arise from different isospin interaction terms in (\ref{mmDM3}) and (\ref{all34}), each mass associated essentially to one of the interactions.

Because of the DM dominance one may say that a single tetrahedron of isospin vectors is a `DM isomagnet'. It is actually a {\it frustrated} DM isomagnet, because the DM interaction between two single tetrons would prefer an angle of 90 degrees between their respective isospin vectors instead of the tetrahedral 'star' configuration formed by the 4 isospin vectors\cite{frust} inside the internal tetrahedron, where the minimum DM energy has to be counterbalanced by geometry.

There is, however, a different interpretation arising from (\ref{mmDM3}) and (\ref{all34}), where one attains attraction among isospins instead of frustration, and furthermore both inner- and inter-tetrahedral interactions turn out to be of order $v_F$.
Namely, there is a Hamiltonian for the interaction between 2 isospins $\vec Q_{i}$ and $\vec Q_{j}$ with minimum energy at the tetrahedral angle $\theta_{tet}=\arccos(-\frac{1}{3})$, thus stabilizing the tetrahedral 'star' arrangement. As compared to (\ref{mm3}) and (\ref{mm3444}) this Hamiltonian has the form
\begin{eqnarray} 
H \sim \sum_{i\neq j=1}^4  \vec Q_{i} \vec Q_{j} - \frac{3}{2} \sum_{i\neq j=1}^4  (\vec Q_{i} \times \vec Q_{j} ) ^2
\label{all30}
\end{eqnarray}
Since the Heisenberg term is $\sim \cos(\theta)$ and the DM-term involves $\sin(\theta)$, their linear combination (\ref{all30}) can be shown to have a minimum at  $\theta_{tet}$.
One can then rewrite the top ($K_{LL}$) and charm ($J_{LL}$) mass part of the Hamiltonian $H_H+H_D$ eqs. (\ref{mmDM3}) and (\ref{all34}) as a sum of 2 contributions
\begin{eqnarray} 
\frac{2}{3} K_{LL}  [  \sum_{i\neq j=1}^4 \vec Q_{Li} \vec Q_{Lj}  -\frac{3}{2} \sum_{i\neq j=1}^4   (\vec Q_{Li} \times \vec Q_{Lj} ) ^2 \,]
- (\frac{2}{3}K_{LL}+J_{LL})\sum_{i\neq j=1}^4 \vec Q_{Li} \vec Q_{Lj} 
\label{all37}
\end{eqnarray}
where the first term is assumed to arise from the inner tetrahedral interactions,
and the second from the inter ones. Both the inner and inter contributions now are of order $v_F (\sim K_{LL})$, the inner having a minimum at $\theta_{tet}$ thus stabilizing any tetrahedron of isospins, and the inter with coupling $J:=\frac{2}{3}K_{LL}+J_{LL}$ being a `ferromagnetic' Heisenberg interaction which supports the alignment of any 2 neighboring tetrahedrons of isospins.

\begin{center}
{\bf V. Isospin Conserving Torsion and the Masses of the First Family}
\end{center}

In the previous sections it was shown how the heaviness of the third family is related to large DM couplings. Afterwards masses of the quarks and leptons of the second family were obtained from Heisenberg exchange. In this section it will be seen how the small masses of the first family can be obtained from isospin conserving torsional interactions.

It turns out that torsional interactions give contributions to the masses of all families. 
However, since they are assumed to be small, the 2 heavy families remain 
dominated by DM and Heisenberg couplings, as given in (\ref{all36}).

The structure of torsional interactions is quite simple. They correspond to a generalization of Hooke's law to rotations, where instead of an exerting force which is proportional to the stretch x there is an exerting torque which is proportional to the stretch angle $\varphi$. 
\begin{eqnarray}
I \frac{d^2\varphi}{dt^2}   = -C_T^2 \varphi
\label{hoo1}
\end{eqnarray}
with some constant $C_T$. The energy of the system is given by 
\begin{eqnarray}
E_T = \frac{1}{2} I (\frac{d\varphi}{dt})^2+ \frac{1}{2} C_T^2  \varphi^2
\label{hoo10}
\end{eqnarray}
 with I the moment of inertia.

By differentiation one can see that the second order differential equation (\ref{hoo1}) is equivalent to $d\varphi/dt=iC_T\varphi$ and thus to the first order equation
\begin{eqnarray}
\frac{dQ}{dt} = i C_T Q  
\label{hoo2}
\end{eqnarray}
where $Q=I d\varphi/dt$ is the angular momentum and $dQ/dt$ the torque.
 
In the present context (\ref{hoo2}) is more suitable than (\ref{hoo1}), because it can be immediately added to the system of differential equations for the $\vec Q_{Li}$ and $\vec Q_{Ri}$ which was obtained in (\ref{allnxxg}) and (\ref{allnxxg1}) for the DM and Heisenberg interactions. Using the notation introduced in (\ref{anx5}) one has
\begin{eqnarray}
\frac{d\vec\delta_{Li}}{dt}=i C_{LL} \vec \Delta_{LLi}   + i C_{LR} \vec \Delta_{LRi} 
\label{v377g}\\
\frac{d\vec\delta_{Ri}}{dt}=i C_{LR}  \vec\Delta_{RLi}   + i C_{RR}  \vec\Delta_{RRi} 
\label{v377}
\end{eqnarray}
where the couplings $C_{LL}$, $C_{LR}$ and $C_{RR}$ generalize $C_T$ to $\vec Q_L$ and $\vec Q_R$. 

In the formulation (\ref{v377}) care has been taken to maintain isospin conservation as defined in (\ref{tm31}). This requirement leads to the appearance of the linear combinations $\Delta$ given in (\ref{anx5}). 

Since (\ref{v377g}) and (\ref{v377}) give the only mass contributions to the first family, the C-couplings can be chosen to accommodate the mass of the up quark, down quark and electron, respectively. Namely, one arrives at the mass formulas
\begin{eqnarray}
m_e&=&6 C_{LR}\\
m_u&=&-2C_{LL}+3C_{LR}+2C_{RR}-W_C\\
m_d&=&-2C_{LL}+3C_{LR}+2C_{RR}+W_C
\label{v37a7}
\end{eqnarray}
where 
\begin{eqnarray} 
W_C:= \sqrt{4(C_{LL}+C_{RR})^2+C_{LR}^2}
\label{allep36}
\end{eqnarray}
Then, using the phenomenological values
\begin{eqnarray}
m_e=0.51\; MeV \qquad \quad m_u=1.7 \;MeV \qquad \quad m_d=4.7\; MeV 
\label{v37b7}
\end{eqnarray}
one obtains
\begin{eqnarray}
C_{LR}=0.085\; MeV \qquad \quad  C_{LL}=1.13 \;MeV \qquad \quad C_{RR}=0.49 MeV
\label{v37c7}
\end{eqnarray}

\begin{center}
{\bf VI. Neutrino Masses and Isospin Nonconservation}
\end{center}

In discussions of neutrino masses there is always the question whether they are of Dirac or Majorana type. Within the tetron model, neutrinos have the same spacetime properties as the other quarks and leptons, because all isospin excitations inherit their SO(3,1) transformation properties from the underlying octonion representation of SO(6,1) - which is Dirac.

This means, neutrinos are special only because of their small masses. In the tetron model small neutrino masses arise in the following way: among the 24 isospin excitations, which are the quarks and leptons, there are always 3 $G_4$-singlet modes which are approximately massless. This has to do with the conservation of total isospin. The 3 masses are suppressed because they correspond to the vibrations of the 3 components of the total internal angular momentum vector within one tetrahedron
\begin{equation} 
\vec \Sigma :=\sum_{i=1}^4  (\vec Q_{Li}+\vec Q_{Ri}) =\sum_{i=1}^4 \vec Q_i =\frac{1}{2} \sum_{i=1}^4 \Psi_i^\dagger \vec \tau\Psi_i
\label{tm32}
\end{equation}
Whenever this quantity is conserved
\begin{equation} 
d\vec \Sigma/dt =0 
\label{tm31}
\end{equation}
the neutrino masses will strictly vanish. In fact, the combinations of Heisenberg, DM and torsional interactions (\ref{mmDM3}), (\ref{all34}), (\ref{v377g}) and (\ref{v377}) considered so far, conserve total isospin. They fulfill (\ref{tm31}) and give no contribution to the neutrino masses. A signal for the conservation of isospin is the appearance in all those equations of the linear combinations 
\begin{equation} 
\vec \Delta_i=- 3 \vec \delta_i + \sum_{j\neq i} \vec \delta_j 
\label{tm37f}
\end{equation}
The $\Delta_i$ enter $d\vec \Sigma$  in the form of the sum $\sum_i \vec \Delta_i$ - and this sum trivially vanishes.

Nonvanishing contributions to the neutrino masses will be derived below in a systematic and comprehensive way. In order to enlighten the procedure, first consider as a simple example an isospin conserving torque of the form
\begin{eqnarray} 
\frac{d\vec Q_1}{dt} \sim (\vec Q_2-\vec Q_1) +(\vec Q_3-\vec Q_1) +(\vec Q_4-\vec Q_1)=\vec \Delta_1
\label{alln00}
\end{eqnarray}
and compare it with an isospin violating one
\begin{eqnarray} 
\frac{d\vec Q_1}{dt}= i N_T (\vec Q_1-\vec Q_0) =N_T \vec \delta_1
\label{alln1}
\end{eqnarray}
with some tiny new coupling $N_T$ and $\vec Q_0=\langle \vec Q_1\rangle$ denoting the ground state value of $\vec Q_1$. Similarly $dQ_j/dt=iN_T (\vec Q_j-\langle \vec Q_j\rangle)$ for $j=2,3,4$.

What is the physical meaning of such an isospin violating contribution? After all, (\ref{alln1}) does not exhibit any interaction of $\vec Q_1$ with the other $\vec Q_{2,3,4}$. It is an isospin non conserving reset torque towards $\vec Q_0$ and effects a mysterious steady gain or loss of isospin, which certainly needs understanding. 

In my opinion there is only one plausible explanation: in order that isospin does not disappear into nirvana, the most straightforward assumption is the existence of some kind of nucleus sitting at the center of each tetrahedron and to which isospin can be transferred, at least in tiny doses. There may be other explanations, but I find this one particularly appealing, because one may speculate that the nuclei are responsible for an additional stabilization of the substrate's skeleton structure in Fig. 1. 

As seen below, in addition to giving neutrino masses, the coupling $N_T$ also enters all the other quark and lepton mass formulas. Therefore, there is {\it always} this tiny exchange of isospins with the nucleus, whenever a tetrahedron of isospins gets excited to a quark or a lepton.

With contributions (\ref{alln1}) alone, all 3 neutrinos $\nu_e$, $\nu_\mu$ and $\nu_\tau$ get the same mass of order $N_T$. To obtain different masses it is instructive to remember how the different masses for the 3 families were obtained in the case of the other quarks and leptons, namely by use of isospin-preserving Heisenberg, torsional and DM interactions. Analogously, one may construct isospin violating DM and Heisenberg interactions by replacing $\Delta\rightarrow\delta$ in (\ref{allnxxg}) and (\ref{allnxxg1}). One obtains
\begin{eqnarray} 
\frac{d\vec Q_{Li}}{dt}
&=& i N_T (\vec Q_{Li}- \vec Q_0) + N_H (\vec Q_{Li}\times \vec Q_0)
\\ & & +2N_{D} \{- (\vec Q_{Li}\times \vec Q_0) + i  (- (\vec Q_{Li}- \vec Q_0) +  ((\vec Q_{Li}- \vec Q_0)\vec Q_0)\vec Q_0]\} \nonumber
\\ &=& i N_{T} \vec \delta_{Li}+ N_{H} (\vec \delta_{Li}\times \vec Q_0) \nonumber
+2N_{D} \{- ( \vec\delta_{Li} \times \vec Q_0)+ i  (- \vec \delta_{Li} + ((\vec \delta_{Li}\vec Q_0)\vec Q_0)]\}
\label{alln2g}
\end{eqnarray}
and similarly for $\vec Q_{Ri}$. This procedure leads to different masses for the 3 neutrino mass eigenstates $\nu_1$, $\nu_2$ and $\nu_3$ of the following form 
\begin{eqnarray} 
m(\nu_1)=4 N_T \qquad m(\nu_2)=N_T+N_H \qquad m(\nu_3)= N_H +4 N_{D} - 4N_T 
\label{alln3}
\end{eqnarray}
As mentioned before, all other quarks and leptons get similar contributions to their masses from $N_T$, $N_H$ and $N_{D}$. However, since the isospin violating couplings $N$ are assumed to be tiny ($\le 1$eV), they can be neglected in the mass formulas which were presented in the preceeding sections. 

One may accommodate (\ref{alln3}) to the results from neutrino oscillation experiments. Consider first the case of the so called `normal mass hierarchy' $m(\nu_1) < m(\nu_2) \ll m(\nu_3)$ where
\begin{eqnarray} 
m(\nu_1)/eV=0.001  \qquad m(\nu_2)/eV=0.0087 \qquad m(\nu_3)/eV=0.048 
\label{alln41}
\end{eqnarray}
Lacking experimental informations on $m(\nu_1)$ I have guessed here a value of $0.001$ eV. In the normal hierarchy limit $m(\nu_1) \ll m(\nu_2) \ll m(\nu_3)$ one sees that $m(\nu_1)$ is a measure of the torsional coupling $N_{T}$,  $m(\nu_2)$ measures the strength $N_{H}$ of the Heisenberg coupling and $m(\nu_3)$ of the DM coupling $N_{D}$. The situation is thus similar as for the other quarks and leptons, where the heavy family mass is dominated by DM interactions, the second family by Heisenberg and the light family by torsional couplings.  

In the case of the so called `inverted hierarchy' one has 
\begin{eqnarray} 
m(\nu_1)/eV\approx  m(\nu_2)/eV=0.0245 \qquad m(\nu_3)/eV=0.001 
\label{alln42}
\end{eqnarray} 
where this time the assumption is made on the (unknown) mass $m(\nu_3)$. Trying to accommodate (\ref{alln42}) with (\ref{alln3}) one obtains $N_H\approx 0$. 
At the same time a small $m(\nu_1)$ leads to $N_D\approx N_T$, i.e. an accidental compensation between torsional und DM contribution is needed to occur. 
 
In summary, it was found in this section, that the masses $m(\nu_1)$, $m(\nu_2)$ and $m(\nu_3)$ are a measure of the strength of the isospin-violating torsional, Heisenberg and DM interactions, respectively. This happens in a similar way, as the masses of the first, second and third family of quarks and charged leptons are determined by the strength of the isospin-conserving torsional, Heisenberg and DM interactions, cf. the discussion at the beginning of Sect. V.

\begin{center}
{\bf VII. Lepton Eigenstates}
\end{center}

In the previous sections the focus of discussion has been laid on the eigenvalues of the system, i.e. on quark and lepton masses. In the course of the calculations the transition from isospin to mass eigenstates has been carried out via an appropriate diagonalization process and has led to numerical values for the quark and lepton masses. 

Actually, the dynamic equations for the isospin vectors allow to calculate the eigenfunctions as well. Namely, the Mathematica program presented in Appendix A gives the physical mass eigenstates in terms of the isospin eigenstates, provided one simply changes the command 'eigenvalues' to 'eigensystem' in the last line of the code. 

While the masses correspond to the eigenvalues, CKM and PMNS mixings can be inferred from the eigenvectors. Details of this deduction are presented in Sect. IX. Here I will concentrate on explicitly representing the quark and lepton mass states in terms of the isospin eigenstate vectors. To that end, the following definitions are used: 
\begin{eqnarray} 
\ket{\vec S}:= \vec\delta_L  \qquad\qquad\qquad \ket{\vec T}:= \vec\delta_R 
\label{nn77}
\end{eqnarray}
Dirac's notation with bra and ket states is applied to make the mechanism more transparent. The index $i=1-4$ counting the tetrahedral sites is left out for reasons discussed below. 

The quantities (\ref{nn77}) are orthonormal vector states and can be used to write down the equations for the neutrino mass eigenstates, as obtained from Appendix A:
\begin{eqnarray} 
\ket{\nu_{e,m}}&=&\frac{1}{\sqrt{6}} [(\ket{S_x} +\ket{T_x}) +  (\ket{S_y} + \ket{T_y})  + (\ket{S_z} + \ket{T_z}) ] \nonumber \\
\ket{\nu_{\mu, m}}&=&\frac{1}{\sqrt{6}} [ (\ket{S_x} + \ket{T_x}) +\omega (\ket{S_y} + \ket{T_y}) + \bar \omega (\ket{S_z} + \ket{T_z})] \nonumber \\
\ket{\nu_{\tau, m}}&=& \frac{1}{\sqrt{6}} [(\ket{S_x} +\ket{T_x}) +\bar\omega (\ket{S_y} + \ket{T_y}) + \omega  (\ket{S_z} + \ket{T_z}) ] 
\label{allnn5}
\end{eqnarray}
The corresponding result for the charged leptons is
\begin{eqnarray} 
\ket{e_m}&=&\frac{1}{\sqrt{6}} [  (\ket{T_x}-\ket{S_x}) +  (\ket{T_y}-\ket{S_y})  + (\ket{T_z}-\ket{S_z}) ] \nonumber \\
\ket{\mu_m}&=& \frac{1}{\sqrt{6}}[(\ket{T_x}-\ket{S_x}) + \omega (\ket{T_y}-\ket{S_y})  + \bar \omega (\ket{T_z}-\ket{S_z}) ] \nonumber  \\
\ket{\tau_m}&=& \frac{1}{\sqrt{6}}[(\ket{T_x}-\ket{S_x}) + \bar \omega (\ket{T_y}-\ket{S_y})  +\omega (\ket{T_z}-\ket{S_z}) ] 
\label{allee5}
\end{eqnarray}
Reference is made to the isospin of only 1 of the 4 tetrons within a tetrahedron, because the contributions from the other 3 tetrons to the eigenstates are identical, and for simplicity not included. 


The appearance of the complex numbers
\begin{eqnarray} 
\omega=-\frac{1-i\sqrt{3}}{2} \qquad\qquad\qquad \bar\omega=-\frac{1+i\sqrt{3}}{2}
\label{mack151}
\end{eqnarray}
corresponding to rotations by 120 and 240 degrees are an effect of the underlying tetrahedral symmetry. They turn the expressions (\ref{allnn5}) and (\ref{allee5}) into $G_4$-symmetry adapted functions.

The lepton mass states actually can be brought to the much more compact form 
\begin{eqnarray} 
\begin{bmatrix}  \ket{\nu_{e m}}\\     \ket{\nu_{\mu m} }   \\  \ket{\nu_{\tau m} }    \\ \end{bmatrix}
=Z\begin{bmatrix}  \ket{V_x}\\     \ket{V_y}   \\     \ket{V_z}    \\ \end{bmatrix}
\qquad\qquad\qquad\qquad
\begin{bmatrix}  \ket{e_m}\\     \ket{\mu_m}   \\     \ket{\tau_m}    \\ \end{bmatrix}
=Z\begin{bmatrix}  \ket{A_x}\\     \ket{A_y}   \\     \ket{A_z}    \\ \end{bmatrix}
\label{fo77}
\end{eqnarray}
by using the definitions
\begin{eqnarray} 
\ket{\vec V}= \frac{1}{\sqrt{2}}(\ket{\vec S}+\ket{\vec T})  \qquad \qquad\qquad
\ket{\vec A}= \frac{1}{\sqrt{2}}(\ket{\vec T}-\ket{\vec S}) 
\label{nn77a}
\end{eqnarray}
and the $Z_3$ Fourier transform matrices
\begin{eqnarray} 
Z=\frac{1}{\sqrt{3}}
\begin{bmatrix}
   1 & 1 & 1  \\
    1 & \omega & \bar\omega   \\
    1 &  \bar\omega &  \omega   \\
\end{bmatrix}
\qquad \qquad 
Z^\dagger=\frac{1}{\sqrt{3}}
\begin{bmatrix}
   1 & 1 & 1  \\
  1 &  \bar\omega & \omega   \\
   1 &   \omega &  \bar\omega   \\
\end{bmatrix}
\label{matz35}
\end{eqnarray}
Note in passing $\ket{\vec V}$ and $\ket{\vec A}$ describe fluctuations of the isospin vectors and axial vectors $\Psi^\dagger \vec \tau\Psi$ and $\Psi^\dagger \vec \tau \gamma_5 \Psi$ around their ground state values.

My calculations show that the eigenfunctions (\ref{allnn5}), (\ref{allee5}) and (\ref{fo77}) are stable against variations of all the isospin couplings introduced in the last chapters. As shown in Sect. IX, this implies that the neutrino mixing matrix does not depend on any fermion mass values and leads to a stable and unambiguous prediction for the PMNS matrix. 

\begin{center}
{\bf VIII. Quark Eigenstates}
\end{center}

The Mathematica output from Appendix A for the quark mass states in terms of the isospin eigenstates (\ref{nn77}) at first sight looks rather cumbersome, but can be simplified for several reasons. First of all, on grounds of symmetry it is not necessary to write down the full 24$\times$24 output. As discussed in the last section, for the case of leptons all 4 tetrons I, II, III and IV on the tetrahedron contribute in the same way, i.e. the structure of the eigenstate is always of the form of a sum I+II+III+IV, so that for the presentation it sufficed to write down the contribution from tetron I. Similarly, the quark states have a 
recurring form $3\times$I-II-III-IV (for one color, and $3\times$II-I-III-IV and $3\times$III-I-II-IV for 
the other two). Knowing this, it is enough to present the contribution of one of the tetrons to one of the colors\footnote{The quark triplets are triplets under the Shubnikov group initially. For the question how to interpret them as SU(3) color triplets one may consult the appendix in \cite{bodotalk}.}.

A complication arises from the fact that Mathematica cannot distinguish between degenerate eigenstates. Therefore, in order to determine for each quark flavor the state of a definite color (e.g. the one of the form $3\times$I-II-III-IV) I had to introduce artificially a tiny color breaking coupling in the program Appendix A. My choice was an additional contribution 0.00001*del1u and 0.00001*eel1u to the terms zx5 and zx1 for the isospins of tetron I. Using this trick the program in Appendix A finally leads to the mass eigenstates for the up-type quarks 
\begin{eqnarray} 
\ket{u_m}&=&\frac{1}{\sqrt{3}\sqrt{1+\epsilon_{1}^2}}  [(\ket{S_x} +\epsilon_{1} \ket{T_x}) +  (\ket{S_y} + \epsilon_{1} \ket{T_y})  + (\ket{S_z} + \epsilon_{1} \ket{T_z}) ] 
\nonumber \\
\ket{c_m}&=& \frac{1}{\sqrt{3}\sqrt{1+\epsilon_{2}^2}} [ (\ket{S_x} +\epsilon_2 \ket{T_x}) +\omega (\ket{S_y} + \epsilon_2 \ket{T_y}) + \bar \omega (\ket{S_z} + \epsilon_2 \ket{T_z})] \nonumber \\
\ket{t_m}&=& \frac{1}{\sqrt{3}\sqrt{1+\epsilon_{3}^2}} [(\ket{S_x} +\epsilon_{3} \ket{T_x}) +\bar\omega (\ket{S_y} + \epsilon_{3} \ket{T_y}) + \omega  (\ket{S_z} + \epsilon_{3} \ket{T_z}) ]
\label{allup}
\end{eqnarray}
and for the down quarks
\begin{eqnarray} 
\ket{d_m}&=&\frac{1}{\sqrt{3}\sqrt{1+\epsilon_{1}^2}} [  (\ket{T_x}-\epsilon_1 \ket{S_x}) +  (\ket{T_y}-\epsilon_1 \ket{S_y})  + (\ket{T_z}-\epsilon_1 \ket{S_z}) ] \nonumber \\
\ket{s_m}&=& \frac{1}{\sqrt{3}\sqrt{1+\epsilon_{2}^2}} [(\ket{T_x}-\epsilon_2 \ket{S_x}) + \omega (\ket{T_y}-\epsilon_2 \ket{S_y})  + \bar \omega (\ket{T_z}-\epsilon_2 \ket{S_z}) ] \nonumber  \\
\ket{b_m}&=& \frac{1}{\sqrt{3}\sqrt{1+\epsilon_{3}^2}} [(\ket{T_x}-\epsilon_3 \ket{S_x}) + \bar \omega (\ket{T_y}-\epsilon_3 \ket{S_y})  +\omega (\ket{T_z}-\epsilon_3 \ket{S_z}) ]
\label{alldown}
\end{eqnarray}
As discussed before, reference is made to the isospins $\vec S$ and $\vec T$ of tetron I only\footnote{Formally, the lepton eigenfunctions (\ref{allnn5}) and (\ref{allee5}) are recovered by chosing $\epsilon_{3} = \epsilon_{2} = \epsilon_{1} = 1$. It should be stressed, however, that this is only formally true, because the quark states (\ref{allup}) and (\ref{alldown}) are defined in a different space than the lepton states; see the above discussion on eigenfunctions I+II+III+IV for leptons and 3$\times$I-II-III-IV for quarks.}.

Three coefficients $\epsilon_{1,2,3}$ appear in these equations. They depend on the DM, HH and torsional isospin couplings introduced in Sects. IV, V and VI and can be calculated within the model (numbers will be given below). Since there is a one-to-one relation between these couplings and the Yukawa couplings and fermion masses, the $\epsilon_{i}$ may be considered to depend on the quark (and lepton) masses. Qualitatively, this dependence is such, that variation of the i-th family masses modifies $\epsilon_{i}$ only, and hardly the other $\epsilon_{j}$. Furthermore, $\epsilon_{i}$ goes up when increasing the charged lepton mass of family i, while it goes down with increasing quark masses of family i. It is worthwhile to stress that in any case there is an appreciable dependence on the LR-couplings, which determine the charged lepton masses.  

In the case of the light familiy $\epsilon_1$ as well as the fermion masses are determined solely by the isospin conserving torsional couplings (\ref{v377g}), (\ref{v377}) and (\ref{v37c7}). Masses are given in (\ref{v37a7}) and the $\epsilon_1$ parameter is
\begin{eqnarray} 
\epsilon_1 = \frac{8 C_{LL}^3 - 3C_{LL}C_{LR}^2+8C_{LL}^2C_{RR}-5C_{LR}^2C_{RR}
+(4C_{LL}^2+2C_{LR}^2) W_C}
{C_{LR} [-2C_{LL}^2+2C_{LR}^2-2C_{RR}^2 + (C_{LL}+C_{RR}) W_C]   } 
\label{allep35}
\end{eqnarray}
With a little algebra one can rewrite this formula so that $\epsilon_1$ depends only on the masses $M_{U1}=m_u$, $M_{D1}=m_d$ and $M_{L1}=m_e$ of the up quark, the down quark and the electron
\begin{eqnarray} 
\epsilon_{1} = \frac{2 f_+ (f_+ +f_-)^2 +f_0^2 (f_- -4 f_+) 
+ [2 f_0^2- (f_+ +f_-)^2]  \sqrt{4 f_+^2 +f_0^2}  }
{ f_0 [f_- \sqrt{4 f_+^2 +f_0^2} +2f_0^2-f_+^2 - f_-^2]}
\label{sm15ex}
\end{eqnarray}
where I have introduced the abbreviations 
\begin{eqnarray} 
f_+ =\frac{1}{4}\sqrt{(M_{U1} + M_{D1})^2-\frac{2}{3} M_{L1}^2} \quad
f_-=\frac{1}{4} (M_{U1}-M_{D1}+M_{L1}) \quad f_0=\frac{M_{L1}}{6}
\label{sm16ex}
\end{eqnarray}
This complicated result can be approximated to very good precision by
\begin{eqnarray} 
\epsilon_{1} = \frac{1}{6} \frac{M_{L1}}{M_{U1}+M_{D1}}
\label{sm14ex}
\end{eqnarray}
Interestingly, it turns out that the corresponding results for $\epsilon_2$ and $\epsilon_3$ are formally identical to (\ref{sm14ex}), except that one has to replace the values for the up, down and electron mass by the corresponding mass values of the second and the third family. 
\begin{eqnarray} 
\epsilon_{i} = \frac{1}{6} \frac{M_{Li}}{M_{Ui}+M_{Di}}
\label{sm14exxx}
\end{eqnarray}
In other words, one obtains the result for the eigenstates irrespective of what kind of coupling (DM, Heisenberg or torsion) is considered.

Another interesting point is that there is the dependence of (\ref{sm14exxx}) on the lepton masses. Since the $\epsilon_i$ enter the CKM matrix, a lepton mass dependence appears in the CKM elements, cf. (\ref{wck41}) and (\ref{ckm1aa}).

Using the plain quark and lepton mass values given by the particle data group\cite{pdg} the formula (\ref{sm14exxx}) yields 
\begin{eqnarray} 
\epsilon_{1} = 0.0140  \qquad \epsilon_{2} =0.0128 \qquad \epsilon_{3} =0.00171
\label{sm14}
\end{eqnarray}
It will be argued later in (\ref{sm15}) that instead of the low energy values (\ref{sm14}) $\epsilon_{i}$-values near the Planck scale should better be considered. They show a more pronounced hierarchy $\epsilon_{1} \gg  \epsilon_{2} \gg\epsilon_{3}$ than (\ref{sm14}).

Instead of $\epsilon_{1,2,3}$ one may introduce three angles $\alpha_{1,2,3}$ to describe the rotations between S and T states implicit in the eigenfunctions (\ref{allup}) and (\ref{alldown}), with $\alpha_i:=\arctan (\epsilon_{i})$.
The $\alpha_i$ turn out to be related to the angles appearing in the standard parametrization of the CKM matrix.

\begin{center}
{\bf IX. Quark and Lepton Mixing Matrices}
\end{center}

Experiments show that there is a mixing between the flavor and mass eigenstates of the 3 neutrino species, and this can be described by a unitary matrix, the PMNS neutrino mixing matrix\cite{maki,giganti}. The experimentally relevant quantities are the absolute values of the matrix elements, which describe the amount of admixture of the flavor into mass eigenstates, and the leptonic Jarlskog invariant which describes any possible CP violation in the leptonic sector. 

Since the discovery of neutrino oscillations, many models of neutrino mass and mixings have been constructed. The most straightforward approach is to incorporate Dirac neutrino masses into the Standard Model by introducing three right-handed neutrinos coupled to a Higgs field analogously to the quarks and charged leptons - but unfortunately, within the SM the values of the mixing parameters cannot be predicted. 

{\bf Leading symmetric Approximation}

In a first step a leading order result for the mixing matrix will be derived which is
\begin{eqnarray} 
V_{PMNS}&=&\exp \Biggl\{\frac{i}{{\sqrt{3}}} \begin{bmatrix}
   0 & 1 & 0  \\
    1 & 1 & -1  \\
     0 &  -1 & -1  \\
\end{bmatrix}\Biggr\} \nonumber \\
&=& \begin{bmatrix}
   0.8467-i 0.0300 & -0.1489+i 0.4861 & 0.1532-i 0.00051  \\
    -0.1489-i 0.4861 & 0.5446+i 0.4568 & -0.00433 - i 0.4858  \\
      0.1532-i 0.00051 &  -0.00433 - i 0.4858 & 0.6892-i 0.5153  \\
\end{bmatrix}
\label{pmn0a}
\end{eqnarray}
while an improved formula will be given later in (\ref{p0imp}). 

The leading order expression (\ref{pmn0a}) is a complex, symmetric and unitary matrix, and the absolute values of the matrix elements can be calculated numerically and compared to measurements 
\begin{eqnarray} 
\begin{bmatrix}
   0.843 & 0.510 & 0.153  \\
    0.510 & 0.711 & 0.486  \\
      0.153 &  0.486 & 0.861  \\
\end{bmatrix}
\quad vs. \quad
\begin{bmatrix}
   0.80-0.85 & 0.51-0.58 & 0.142-0.155  \\
    0.23-0.51 & 0.46-0.69 & 0.63-0.78  \\
      0.25-0.53 &  0.47-0.70 & 0.61-0.76  \\
\end{bmatrix}
\label{pmn1a}
\end{eqnarray} 
By inspection one concludes that the agreement is reasonable but not optimal, with the 23 entry being the most critical. 
The first row, which is best measured, is also best fitting. Concerning the other rows, the experimental results in (\ref{pmn1a}) are non-symmetric, though with very large errors. It will be described later, in connection with (\ref{p0imp}) and (\ref{pmnxxa}), how (\ref{pmn0a}) can be improved by additional non-symmetric contributions so that complete agreement within the errors is obtained.

A prediction for the leptonic Jarlskog invariant\cite{jarls} can be calculated from  (\ref{pmn0a}) as  
\begin{eqnarray} 
J_{PMNS}=\Im (V_{e1} V_{\mu 2} \bar V_{e2} \bar V_{\mu 1} ) =-0.0106  
\label{pmn2a}
\end{eqnarray} 
This value is large as compared to the Jarlskog parameter of the CKM matrix\cite{pdg}. $J_{PMNS}$ has not been measured so far, although there are experimental indications that leptonic CP violation is indeed rather large\cite{jarl2}. 

{\bf Motivation and Proof of (\ref{pmn0a})}

My calculations show that the eigenfunctions (\ref{allnn5}), (\ref{allee5}) and (\ref{fo77}) are stable against variations of all the isospin couplings one may use in the Hamiltonian H in (\ref{txxm32}). In consequence, the neutrino mixing matrix does not depend on any fermion mass values. This implies a stable and unambiguous prediction for the PMNS matrix and is in contrast to the CKM matrix in the quark sector, where a mass dependence shows up. 

As well known, the defining equation for the PMNS matrix is
\begin{eqnarray} 
\begin{bmatrix}  \bra{ \nu_{ew}} & \bra{ \nu_{\mu w}} & \bra{ \nu_{\tau w}} \\ \end{bmatrix} W_\mu^+ 
\begin{bmatrix}  \ket{e_w} \\ \ket{\mu_w}  \\ \ket{\tau_w} \\ \end{bmatrix}
=\begin{bmatrix}  \bra{ \nu_{em}} & \bra{ \nu_{\mu m}} & \bra{ \nu_{\tau m}} \\ \end{bmatrix} W_\mu^+ V_{PMNS}
\begin{bmatrix}  \ket{e_m} \\ \ket{\mu_m}  \\ \ket{\tau_m} \\ \end{bmatrix}
\label{wbwwc}
\end{eqnarray} 
where the index $w$ denotes weak interaction eigenstates, and it is understood that we talk about left handed fields only. The mixing matrix is formally given by
\begin{eqnarray} 
V_{PMNS}=V_N V_L^\dagger = 
\begin{bmatrix}
V_{1e} & V_{1\mu} & V_{1\tau} \\
V_{2e} & V_{2\mu} & V_{2\tau} \\
V_{3e} & V_{3\mu} & V_{3\tau}
\end{bmatrix} 
\label{wb111c}
\end{eqnarray} 
where
\begin{eqnarray} 
V_N=
\begin{bmatrix}
\bra{\nu_{e m}}  \\ \bra{\nu_{\mu m}}\\ \bra{\nu_{\tau m}} \\ 
\end{bmatrix}
\begin{bmatrix}
\ket{\nu_{e w}}  & \ket{\nu_{\mu w}}& \ket{\nu_{\tau w}} \\ 
\end{bmatrix}
\qquad
V_L^\dagger=
\begin{bmatrix}
\bra{e_w}  \\ \bra{\mu_w} \\ \bra{\tau_w} \\ 
\end{bmatrix}
\begin{bmatrix}
\ket{e_m}  & \ket{\mu_m}& \ket{\tau_m} \\ 
\end{bmatrix}
\label{wb222c}
\end{eqnarray} 
Replacing the mass eigenstates by the isospin excitations according to (\ref{fo77}) one obtains
\begin{eqnarray} 
V_{PMNS}=Z 
\Biggl\{
\begin{bmatrix}
\bra{V_x}  \\ \bra{V_y} \\ \bra{V_z} \\ 
\end{bmatrix}
\begin{bmatrix}  \ket{ \nu_{ew}} & \ket{ \nu_{\mu w}} & \ket{ \nu_{\tau w}} \\ \end{bmatrix}
\begin{bmatrix}
\bra{e_w}  \\ \bra{\mu_w} \\ \bra{\tau_w} \\ 
\end{bmatrix}
\begin{bmatrix}
\ket{A_x}  & \ket{A_y}& \ket{A_z} \\ 
\end{bmatrix}
\Biggr\} Z^\dagger
\label{wb333c}
\end{eqnarray} 
By inspection one sees that (\ref{wb333c}) exactly compensates all the matrix transformations in (\ref{wbwwc}) and (\ref{fo77}) so as to maintain lepton universality and keep the weak current diagonal in the weak eigenstates.

The brace in (\ref{wb333c}) comprises a matrix of expectation values of the form \begin{eqnarray} 
Y:=\begin{bmatrix} \bra{V_x}  \\ \bra{V_y} \\ \bra{V_z} \\ \end{bmatrix}
\mathcal{O}
\begin{bmatrix}
\ket{A_x}  & \ket{A_y}& \ket{A_z} \\ 
\end{bmatrix}
\label{wb8c}
\end{eqnarray} 
where the inner product 
\begin{eqnarray} 
\mathcal{O}:= 
\begin{bmatrix}  \ket{ \nu_{ew}} & \ket{ \nu_{\mu w}} & \ket{ \nu_{\tau w}} \\ \end{bmatrix}
\begin{bmatrix}
\bra{e_w}  \\ \bra{\mu_w} \\ \bra{\tau_w} \\ 
\end{bmatrix}
\label{dyop1}
\end{eqnarray} 
is a dyadic 1-dimensional operator which acts between the complex 3-dimensional spaces of charged lepton ($\sim \vec S -\vec T$) and antineutrino ($\sim \vec S +\vec T$) states. One may say that it contains all information about what the charged W-boson does to the lepton fields: it changes isospin, mixes families and so on. Weak SU(2) and tetrahedral symmetry force $\mathcal{O}$ to have the form
\begin{eqnarray} 
\mathcal{O}&=&\ket { S_x} \bra  { T_x} + \ket { S_y} \bra  { T_y} + \ket { S_z} \bra  { T_z} -  \ket { T_x} \bra  { S_x} - \ket { T_y} \bra  { S_y} - \ket { T_z} \bra  { S_z}\nonumber \\
&& +\frac{i}{\sqrt{3}}[\ket { S_y} \bra  { S_z} + \ket { S_z} \bra  { S_y} - \ket { T_y} \bra  { T_z} - \ket { T_z} \bra  { T_y}]\nonumber \\
&& +\frac{i}{\sqrt{3}}[\omega \ket { S_x} \bra  { S_y} + \bar \omega \ket { S_y} \bra  { S_x} - \omega \ket { T_x} \bra  { T_y} -  \bar \omega \ket { T_y} \bra  { T_x}]\nonumber \\
&& +\frac{i}{\sqrt{3}}[ \bar \omega \ket { S_x} \bra  { S_z} + \omega \ket { S_z} \bra  { S_x} -  \bar \omega \ket { T_x} \bra  { T_z} - \omega \ket { T_z} \bra  { T_x}]  
\label{wb1}
\end{eqnarray} 
In order to derive (\ref{wb1}) one has to note that SU(2) invariance allows the appearance of dot products and triple products only. The coefficients of these products are then dictated by the tetrahedral symmetry of the isospin vectors. For example, to derive the triple product coefficients one should remember that the $W^+$-boson is defined in the 3 internal dimensions in an analogous manner as a plus circularly polarized wave in 3 spatial dimensions, namely by means of an `internal polarization' vector $\vec e_+=(\vec e_1 + i \vec e_2)/\sqrt{2}$ which is perpendicular to the axis of quantization, in this case given by $\sim (1,1,1)$.
\begin{eqnarray} 
\vec e_1 = \frac{1}{\sqrt{2}} (0,1,-1) \qquad  \qquad \vec e_2=\frac{1}{\sqrt{6}} (-2,1,1)
\label{pol11}
\end{eqnarray} 
Introducing the vector
\begin{eqnarray} 
\vec \Omega = \frac{1}{\sqrt{3}} (1,\omega,\bar \omega) 
\label{pol11999}
\end{eqnarray} 
allowed contributions to $\mathcal{O}$ are of the triple product form
\begin{eqnarray} 
&& \varepsilon_{ijk}  \frac{1}{\sqrt{2}}(\vec e_1+i \vec e_2)_i \ket { Q_j} \bra  { Q'_k}
= -\frac{i}{\sqrt{3}} \, \vec \Omega (\vec Q \times \vec Q') 
= -\frac{i}{\sqrt{3}} \, [\,\ket { Q'_y} \bra  { Q_z} - \ket { Q'_z} \bra  { Q_y}\nonumber \\
&& \qquad\qquad -\omega (\ket { Q'_x} \bra  { Q_z} - \ket { Q'_z} \bra  { Q_x})
+ \bar \omega (\ket { Q'_x} \bra  { Q_y} - \ket { Q'_y} \bra  { Q_x}\,) \,]
\label{pol12}
\end{eqnarray} 
for the ket and bra states belonging to any 2 internal angular momenta $Q$ and $Q'$. These contributions are anti-hermitean, and care must be taken in the definition of the complex triple product when using complex conjugation in the determination of $\mathcal{O}$. 

Note that $\mathcal{O}$ as given in (\ref{wb1}) is universal in the sense that it depends only on properties of the $\Psi$ field, and therefore will appear in identical form within the quark sector and the calculation of the CKM matrix. This fact reflects the quark lepton universality of the W-boson interactions.

Inserting (\ref{wb1}) into (\ref{wb8c}) one obtains 
\begin{eqnarray} 
Y=\begin{bmatrix} \bra{V_x}  \\ \bra{V_y} \\ \bra{V_z} \\ \end{bmatrix}
\mathcal{O}
\begin{bmatrix}
\ket{A_x}  & \ket{A_y}& \ket{A_z} \\ 
\end{bmatrix}
=I+X
\label{wb8cc}
\end{eqnarray} 
i.e. a sum of a hermitean part (the unit matrix $I$) and an anti-hermitean matrix
\begin{eqnarray} 
X=-\frac{i}{\sqrt{3}}\begin{bmatrix}
   0 & \bar\omega & \omega  \\
    \omega & 0 & 1  \\
     \bar \omega &  1 & 0  \\
\end{bmatrix} 
\label{matz325}
\end{eqnarray}
The invariant structure which gives the unit matrix in (\ref{wb8cc}) is the dot product, while the invariant structure belonging to the anti-hermitean contribution X is the triple product. The unit matrix corresponds to no mixing at all, so the origin of a non-trivial PMNS matrix is to be found solely in the triple product terms (\ref{pol12}). 

The result (\ref{wb8cc}) is anti-hermitian and not unitary, because it represents the leading term in the series exp(X) to enter the unitary PMNS matrix in the following way
\begin{eqnarray} 
V_{PMNS}&=&Z e^{X} Z^\dagger =e^{Z X Z^\dagger}  \nonumber \\
&=& \frac{1}{3}\begin{bmatrix}
   1 & 1 & 1  \\
    1 & \omega & \bar\omega   \\
    1 &  \bar\omega &  \omega   \\
    \end{bmatrix}
\exp \Biggl\{\frac{-i}{{\sqrt{3}}} \begin{bmatrix}
   0 & \bar\omega & \omega  \\
    \omega & 0 & 1  \\
     \bar \omega &  1 & 0  \\
\end{bmatrix}\Biggr\}
\begin{bmatrix}
   1 & 1 & 1  \\
  1 &  \bar\omega & \omega   \\
   1 &   \omega &  \bar\omega   \\
\end{bmatrix}  \nonumber \\
&=& \begin{bmatrix}
   0.8467-i 0.0300 & -0.1489+i 0.4861 & 0.1532-i 0.00051  \\
    -0.1489-i 0.4861 & 0.5446+i 0.4568 & -0.00433 - i 0.4858  \\
      0.1532-i 0.00051 &  -0.00433 - i 0.4858 & 0.6892-i 0.5153  \\
\end{bmatrix}
\label{pmn00a}
\end{eqnarray}
Note that this is identical to what was claimed in (\ref{pmn0a}).

{\bf Improved Formula for the PMNS Matrix}

So far only dot product and triple product terms (\ref{pol12}) have been considered as contributing to the operator (\ref{wb1}) and the PMNS result. Actually, there is a third kind of term that needs consideration. Using $\vec\Omega^2 =0$ and the same normalization as in (\ref{pol12}) it is of the form 
\begin{eqnarray} 
-(\vec \Omega \times \vec Q) \, (\vec \Omega \times \vec Q')= (\vec \Omega \vec Q) \, (\vec \Omega \vec Q')
\label{w8899}
\end{eqnarray} 
As shown in the previous sections, quark and lepton masses are related to torsional, Heisenberg and Dzyaloshinskii isospin interactions of the fundamental tetron $\Psi$ field. Furthermore, these three types of interactions completely fix the structure of the model. 

This fact is reflected in the contributions to the operator $\mathcal{O}$: while the dot products and triple products appearing in (\ref{wb1}) parallel the torsional and Heisenberg interactions, (\ref{w8899}) corresponds to the Dzyaloshinskii Hamiltonian. Working out the products $\ket {Q_i} \bra  {Q'_j}$ arising from (\ref{w8899}), it leads to an additional contribution to (\ref{wb1}) which can be comprised by a matrix
\begin{eqnarray} 
D:= \frac{1}{6}
 \begin{bmatrix}
   1 & \omega & \bar \omega  \\
    \omega & \bar \omega & 1  \\
     \bar \omega & 1 & \omega  \\
\end{bmatrix}
- h.c.
\label{ddma1}
\end{eqnarray}
The role of D for (\ref{w8899}) is analogous to that of X for the triple product term.
Combining the X and D contributions an improved formula for the PMNS matrix is obtained
\begin{eqnarray} 
V_{PMNS}=
\exp \Biggl\{  \frac{1}{2}\begin{bmatrix}
   0 & 0 & 0  \\
    0 & 0 & 1  \\
     0 &  -1 & 0  \\
\end{bmatrix}\Biggr\} 
\,
\exp \Biggl\{\frac{i}{{\sqrt{3}}} \begin{bmatrix}
   0 & 1 & 0  \\
    1 & 1 & -1  \\
     0 &  -1 & -1  \\
\end{bmatrix}\Biggr\} 
\label{p0imp}
\end{eqnarray}
This represents a complex and unitary matrix whose absolute value matrix $|V_{PMNS}|$ is not symmetric, in contrast to (\ref{pmn0a}). Its elements are given by
\begin{eqnarray} 
\begin{bmatrix}
   0.847 & 0.510 & 0.153  \\
    0.430 & 0.505 & 0.748  \\
      0.311 &  0.698 & 0.645  \\
\end{bmatrix}
\quad vs. \quad
\begin{bmatrix}
   0.80-0.85 & 0.51-0.58 & 0.142-0.155  \\
    0.23-0.51 & 0.46-0.69 & 0.63-0.78  \\
      0.25-0.53 &  0.47-0.70 & 0.61-0.76  \\
\end{bmatrix}
\label{pmnxxa}
\end{eqnarray} 
and fit the phenomenological numbers to within one standard error. 

The value of the leptonic Jarlskog invariant now is 
\begin{eqnarray} 
J_{PMNS}=0.0118
\label{pnw33}
\end{eqnarray} 
Thus, while the improvement (\ref{p0imp}) only moderately corrects  the absolute values (\ref{pmn1a}), it strongly modifies the prediction (\ref{pmn2a}) for $J_{PMNS}$ (changing even the sign!). This is because in contrast to the absolute values the Jarlskog invariant is dominated by higher orders of the exponential expansion.  

{\bf Application to Quark Mixing}

Mixing in the quark sector has been known since the time of Cabibbo\cite{cab}. Although the mixing percentages are smaller, it is much better measured than in the lepton sector. On the other hand, concerning theory, the predictions for the CKM mixing elements  in the present model are somewhat more difficult to obtain, though parts of the arguments for leptons can be taken over to the quark sector. The idea is again that the mixing matrix counterbalances the deviation of the mass eigenstates from the weak eigenstates in such a way that the charged current effectively acts diagonal on the isospin operators (\ref{nn77}). The main complication is the appearance of mass dependent factors in the quark eigenstates, see below. 


Within the Standard Model, quark masses arise from Yukawa interactions involving the Higgs field, left-handed\footnote{\color{black} The appearance in the microscopic model of chiral fermions and left handed weak interactions has been discussed detailedly in previous work. In essence it is due to the handedness of the tetrahedrons formed by the isospin vectors in isospin space.} quark doublets, and right-handed down- or up-type quark singlets. The Yukawa couplings are given as complex 3$\times$3 matrices, and the mass terms are obtained when the Higgs field acquires a vacuum expectation value $v$. The physical(=mass) states are attained by diagonalizing the Yukawa matrices $Y^U$ and $Y^D$ using 4 unitary matrices $V^{U,D}_{L,R}$ to obtain diagonal mass matrices for up- and down-type quarks
\begin{eqnarray} 
\frac{v_F}{\sqrt{2}} \, Y^U= V_L^{U\dagger} M^U_{diag} V_R^{U} \qquad \qquad 
\frac{v_F}{\sqrt{2}} \,  Y^D= V_L^{D\dagger} M^D_{diag} V_R^{D}  
\label{dig55}
\end{eqnarray} 
These equations show how the SM Yukawa couplings can be built from the mass eigenvalues and the matrices $V^{U,D}_{L,R}$. One may compare this to the present model where\\ 
-the entries of $M^U_{diag}$ and $M^D_{diag}$ are obtained as excitation energies of the isospin Hamiltonian.\\ 
-the matrices $V^{U,D}_{L}$ can be expressed in terms of the isospin states $\ket{\vec S}$ and $\ket{\vec T}$ as will be seen in the explicit representation (\ref{wckm33}) below.
 
As a result of the diagonalization (\ref{dig55}), the left-handed weak charged-current interactions couple to the physical quarks with couplings involving the CKM matrix which is defined as 
\begin{eqnarray}
V_{CKM}&=&V_{U} V_{D}^\dagger = 
\begin{bmatrix}
V_{ud} & V_{us} & V_{ub} \\
V_{cd} & V_{cs} & V_{cb} \\
V_{td} & V_{ts} & V_{tb}
\end{bmatrix} \nonumber \\
&=& \begin{bmatrix}
\langle u_m | u_w \rangle & \langle u_m | c_w \rangle & \langle u_m | t_w \rangle\\
\langle c_m | u_w \rangle & \langle c_m | c_w \rangle & \langle c_m | t_w \rangle\\
\langle t_m | u_w \rangle & \langle t_m | c_w \rangle & \langle t_m | t_w \rangle
\end{bmatrix}
\begin{bmatrix}
\langle d_w | d_m \rangle & \langle d_w | s_m \rangle & \langle d_w | b_m \rangle\\
\langle s_w | d_m \rangle & \langle s_w | s_m \rangle & \langle s_w | b_m \rangle\\
\langle b_w | d_m \rangle & \langle b_w | s_m \rangle & \langle b_w | b_m \rangle
\end{bmatrix}
\label{mmccm}
\end{eqnarray}
Here I have used Dirac's ket and bra notation as described in the last section. $m$ denotes mass eigenstates (the physical states) and $w$ weak interaction eigenstates. 

Knowing the quark mass eigenstates (64) and (65) one may write down the CKM matrix in an analogous fashion as the PMNS matrix (78) for leptons
\begin{eqnarray} 
V_{CKM}&=&\biggl\{ R Z  
\begin{bmatrix} \bra{S_x}  \\ \bra{S_y} \\ \bra{S_z} \\ \end{bmatrix}
+REZ \begin{bmatrix} \bra{T_x}  \\ \bra{T_y} \\ \bra{T_z} \\ \end{bmatrix} 
\biggr\}
\begin{bmatrix} \ket{ u_w}  & \ket{ c_w}& \ket{ t_w} \\ \end{bmatrix}
\times \nonumber\\
& & \begin{bmatrix} \bra{d_w}  \\ \bra{s_w} \\ \bra{b_w} \\ \end{bmatrix}
\biggl\{ \begin{bmatrix} \ket{T_x}  & \ket{T_y}& \ket{T_z} \\ \end{bmatrix}
 Z^\dagger R - \begin{bmatrix} \ket{S_x}  & \ket{S_y}& \ket{S_z} \\ \end{bmatrix}
Z^\dagger E R \biggr\}
\label{wckm33}
\end{eqnarray} 
where the first line is the matrix $V_L^{U}$ and the second line is $V_L^{D\dagger}$
\color{black} 
and the matrices 
\begin{eqnarray} 
E:=\begin{bmatrix}
   \epsilon_1 & 0 & 0  \\
    0 & \epsilon_2 & 0  \\
    0 &  0 & \epsilon_3  \\
\end{bmatrix}
\qquad\qquad
R:=\begin{bmatrix}
   \frac{1}{\sqrt{1+\epsilon_1^2}} & 0 & 0  \\
    0 & \frac{1}{\sqrt{1+\epsilon_2^2}} & 0  \\
    0 &  0 & \frac{1}{\sqrt{1+\epsilon_3^2}} \\
\end{bmatrix}
\label{abktrua}
\end{eqnarray} 
have been introduced.

Just as in the case of leptons (\ref{dyop1}) there is a 1-dimensional dyadic transformation
\begin{eqnarray} 
\mathcal{O} = 
\begin{bmatrix} \ket{u_w}  & \ket{c_w}& \ket{t_w} \\ \end{bmatrix}
\begin{bmatrix} \bra{d_w}  \\ \bra{s_w} \\ \bra{b_w} \\ \end{bmatrix}
\label{dyop2}
\end{eqnarray} 
which operates between the 3-dimensional spaces of up- and down-type quark states. Due to quark-lepton universality, when expressed in terms of operators $\vec S$ and $\vec T$, the operator $\mathcal{O}$ for quarks must be identical to what was used for leptons in (\ref{wb1}).

Restricting, for a moment, on the dot and triple product contributions (\ref{wb1}) as input, one may then calculate $V_{CKM}$ given in (\ref{wckm33}) to be 
\begin{eqnarray} 
V_{CKM}=I + RZX Z^\dagger E R + REZXZ^\dagger R
\rightarrow \exp \{RZX Z^\dagger E R + REZXZ^\dagger R\}
\label{wck41}
\end{eqnarray} 
where I is the 3$\times$3 unit matrix arising from the dot product terms in (\ref{wb1}). The other terms in (\ref{wck41}) are the anti-hermitian contributions from the triple product in (\ref{pol12}) and (\ref{wb1}). They replace the expression $ZXZ^\dagger$ in (\ref{pmn00a}) for leptons. 

Just as in the case of leptons one may improve on this result by including the contributions from (\ref{w8899}), in order to obtain the desired non-symmetric contributions to $|V_{CKM}|$. The improved formula for the CKM matrix reads
\begin{eqnarray} 
V_{CKM}=\exp \{ 2[RZD Z^\dagger E R - REZD^\dagger Z^\dagger R] \}
\exp \{RZX Z^\dagger E R + REZXZ^\dagger R\}
\label{wck41111}
\end{eqnarray} 
In contrast to X in (\ref{matz325}) the matrix D in (\ref{ddma1}) is not anti-hermitian. This fact has been accounted for in the first exponential factor.

Eq. (\ref{wck41111}) allows to evaluate $|V_{CKM}|$ using appropriate values for the fermion masses entering (\ref{sm14exxx}). It must be noted, however, that the low energy values (\ref{sm14}) of the $\epsilon_{i}$ are not useful in this context. Instead one should use running masses near the Planck scale, because the dynamics generates fermion masses originally at Planck scale distances\footnote{A GUT scale is not present in the model. There is only the Fermi scale, defined as the interaction energy of the isospin vectors, and the Planck scale, defined as the binding energy of the fields $\Psi$\cite{bodohiggs}.}. Unfortunately, the predictions for running masses are not very precise because higher order contributions become appreciable at very large scales. Nevertheless, I am using results from the literature\cite{runnmass,juarez} to determine the $\epsilon_{i}$ at high scales.
\begin{eqnarray} 
\epsilon_{1} = 0.35  \qquad \epsilon_{2} =0.070 \qquad \epsilon_{3} =0.0040
\label{sm15}
\end{eqnarray}
unfortunately with a large theoretical error\cite{juarez}, whose magnitude even is hard to estimate. The numbers are for a 2HDM (2 Higgs doublet model) which is known to be the low-energy limit of the microscopic model\cite{bodohiggs}. They exhibit a family hierarchy which will be seen to induce a corresponding hierarchy in the mixing of the quark families. Actually, as discussed in earlier work\cite{bodomasses}, this is to be expected within the present model due to the large top mass which forces the up- and down-type mass eigenstates to be approximately $\sim\vec S$ and $\sim\vec T$, respectively, in (\ref{allup}) and (\ref{alldown}), much unlike the lepton states which are $\sim\vec S \pm \vec T$ according to (\ref{fo77}).

Just as masses, CKM matrix elements are running, i.e. dependent on the scale paramter $t=\ln \frac{E}{\mu}$ where E is the relevant energy scale and $\mu$ the renormalization scale. The running of the absolute values of the CKM matrix elements has been discussed e.g. in \cite{juarez}. It turns out to be remarkably simple, at least in leading order, because it can be given in terms of one universal function h(t). 
\begin{eqnarray}
|V_{CKM}(t)| \approx  
\begin{bmatrix}
|V_{ud}(0)| & |V_{us}(0)| & \frac{|V_{ub}(0)|}{h(t)} \\
|V_{cd}(0)| & |V_{cs}(0)| & \frac{|V_{cb}(0)|}{h(t)} \\
\frac{|V_{td}(0)|}{h(t)} & \frac{|V_{ts}(0)|}{h(t)} & |V_{tb}|(0)
\end{bmatrix} 
\label{mmccrr1}
\end{eqnarray}
For the Jarlskog invariant one has
\begin{eqnarray}
J_{CKM}(t) \approx \frac{J_{CKM}(0)}{h^2(t)} 
\label{mmccrjr}
\end{eqnarray}
In the 2HDM case $h(t)$ is a moderately varying function. According to \cite{juarez} it increases by about 20\% when going from GeV to Planck scale energies.

Using (\ref{wck41111}) and (\ref{sm15}) I have calculated the CKM elements at high energies and then extrapolated them back to GeV energies according to (\ref{mmccrr1}). I obtain the matrix $|V_{CKM}|$ of absolute values 
\begin{eqnarray} 
\begin{bmatrix}
   0.974 & 0.224 & 0.0035  \\
    0.224 & 0.973 & 0.044  \\
      0.0080 &  0.043 & 0.9991  \\
\end{bmatrix}
vs.
\begin{bmatrix}
   0.9734-0.9740 & 0.2235-0.2251 & 0.00362-0.00402  \\
    0.217-0.225 & 0.969-0.981 & 0.0394-0.0422  \\
      0.0083-0.0088 &  0.0404-0.0424 & 0.985-1.043  \\
\end{bmatrix}
\label{ckm1aa}
\end{eqnarray} 
The numbers look reasonable, as compared to the phenomenological values, and show the correct hierarchy and orders of magnitude. However, the theoretical uncertainty from the scale evolution is large and difficult to estimate, in particular concerning quark mass values near the Planck scale. For example, $\epsilon_{1}$ accommodates the Cabbibo angle correctly, whereas the `23'-matrix elements $|V_{ts}|$ and $|V_{cb}|$ tendencially come out too large, while the `13'-elements $|V_{ub}|$ and $|V_{td}|$ are typically too small. These deviations may seem being just 2$\sigma$ effects, but as stressed before the theoretical error from the quark mass evolution is extremely difficult to handle.

Similarly, concerning the Jarlskog invariant one obtains $J_{CKM}=0.000027$, a bit small when compared to the observed value $J_{CKM}=(3.00+0.15-0.09) \times 10^{-5}$.

\begin{figure}[t]
\centering
  \begin{tabular}{@{}ccc@{}}
    \includegraphics[width=.30\textwidth]{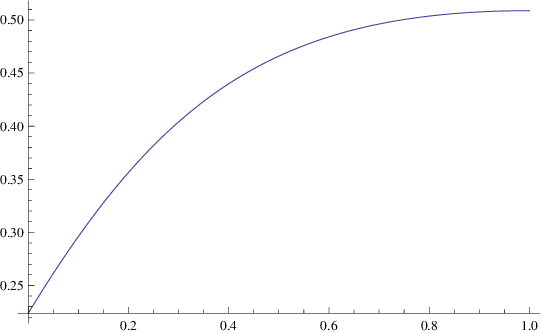} &
    \includegraphics[width=.30\textwidth]{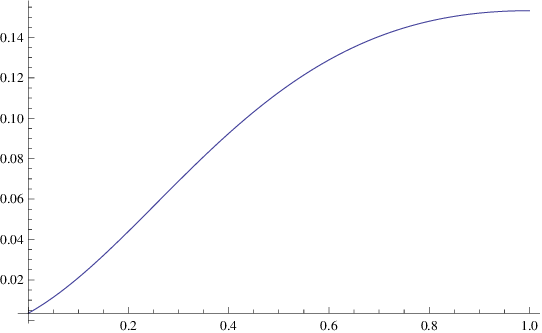} &
    \includegraphics[width=.30\textwidth]{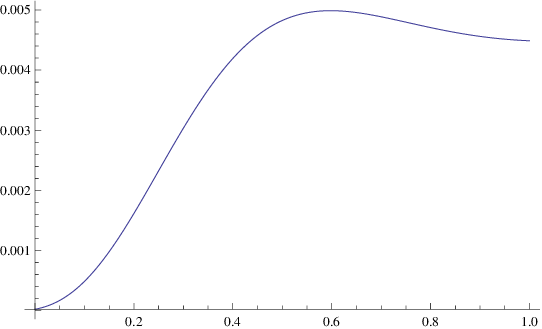}   
  \end{tabular}
\caption{Transition between the CKM and the PMNS limit of the matrix elements $|V_{12}|$ and $|V_{13}|$ and the Jarlskog invariant (from left to right) as a function of the parameter $\alpha$ defined in the main text. For example, $|V_{12}|$ starts with its CKM value 0.224 at $\alpha =0$ and grows towards the PMNS value at $\alpha =1$.}
\end{figure}

In conclusion, explicit analytic and numerical results for the mixing matrices have been presented in this work. Of particular interest are the prediction for the PMNS matrix (\ref{p0imp}) and the fermion mass dependence of the CKM matrix as given by (\ref{wck41111}). Actually, (\ref{wck41111}) is universal in that it embraces (i) the case of no mixing ($\epsilon_{1}=\epsilon_{2}=\epsilon_{3}=0$), (ii) the CKM prediction obtained with $\epsilon_{i}$-values (\ref{sm15}) and (iii) the PMNS formula which formally is given using $\epsilon_{1}=\epsilon_{2}=\epsilon_{3}=1$. To make this visible, I have drawn in Fig. 1 the `12' (i.e. Cabibbo) and the `13' matrix element and the Jarlskog invariant as a function of a parameter $\alpha$. $\alpha$ is introduced to avoid drawing the full $\epsilon_{i}$-dependence of the matrix elements and defined in such a way that it vanishes in the CKM case and takes the value of 1 in the PMNS limit. More precisely, one has 
\begin{eqnarray} 
\epsilon_{1} = 0.35 + 0.65\,\alpha   \qquad \epsilon_{2} =0.07 + 0.93\,\alpha  \qquad \epsilon_{3} =0.004 + 0.996\,\alpha 
\label{sal15}
\end{eqnarray}



\begin{center}
{\bf IX. Summary and Discussion}
\end{center}


This work has shown in detail how the observed spectrum of quarks and leptons can be related to isospin interactions among tetrons. After clarifying the connection between SM Yukawa couplings and the isomagnetic couplings, the magnitudes of the latter were adapted to the observed mass values. Furthermore, in Sect. III it was explained how these coupling parameters themselves can be calculated from exchange integrals involving the fundamental scalar and triplet potentials $V_1$ and $V_3$ among tetrons. The resulting optimized predictions for the masses can be found at the end of the Mathematica program in Appendix A. Numbers are understood in GeV. 

As turns out, Dzyaloshinskii-Moriya couplings are the largest, while Heisenberg interaction terms are smaller. It is a feature of the DM interaction to give masses only to the third family. In particular the top mass is the only excitation with mass of order $v_F$, because it corresponds to a minimum energy of the tetrahedral isospin Hamiltonian (\ref{all30}). This is linked to the SSB of the ordered isospins, i.e. to how the aligned tetrahedral 'stars' are oriented collectively in internal space, thus breaking weak isospin SU(2) symmetry. The ordering takes place below the transition temperature $v_F$, while isospins are distributed randomly (and thus SU(2) symmetric) at temperatures above the Fermi scale.

All other quark and lepton masses naturally turn out to be much smaller than $m_t$. For example, the Heisenberg interactions characteristically give equal contributions to the masses of the second and third family, keeping the first family massless. The first family then obtains its masses from still smaller torsional interactions, as explained in Sect. V.


As a byproduct of the calculations, solutions within the microscopic model to several outstanding classical problems of particle physics have appeared:\\
-The hierarchy problem of why the Fermi scale is so small as compared to the Planck scale. In the microscopic model this is due to the smallness of exchange integrals as compared to direct ones, see the discussion after (\ref{eqrt2h}) in Sect. III.\\
-The tinyness of neutrino masses arises from the conservation of tetron isospin. Isospin violating interactions have been introduced in Sect. VI in order to accommodate reasonable neutrino mass values, and their physical origin has been clarified. Note that isospin is not an abstract concept here but corresponds to real rotations in the 3 extra dimensions.

Concerning the observed quark and lepton mixing a detailed analysis will follow in \cite{bodopmns}. As basis for such an examination in the present work all quark and lepton states of the 3 families have been listed as eigenfunctions of the tetron isospin Hamiltonian. In the microscopic model the internal dynamics of quarks and leptons is intertwined, and therefore it is not surprising that the quark states not only depend on the quark but also on the lepton masses. Details of these and other dependencies were discussed in Sect. VIII.

Neutrino eigenstates are given as vibrations $\vec S+\vec T =\delta \vec Q_L+\delta \vec Q_R$, i.e. of the total isospin vector, and the 3 charged leptons as vibrations $\vec S-\vec T = \delta \vec Q_L-\delta \vec Q_R$. As shown in \cite{bodopmns} this corresponds to large mixing in the lepton sector and large values of the PMNS matrix elements\cite{pdg}. On a qualitative level this should be no surprise in view of the discussion of isospin conservation in Sect. VI, because isospin conservation explains the appearance of sums $\vec S + \vec T$ for neutrinos and - for reasons of orthogonality - of differences $\vec S - \vec T$ for the charged leptons. Neutrino mass eigenstates are thus 'far away' from the isospin states $\vec S$ and $\vec T$, and the resulting PMNS mixing matrix will be 'far away' from the unit matrix. This is much in contrast to the case of quarks where the mass eigenstates are small deviations $\vec S+\epsilon \vec T$ and $-\epsilon \vec S +\vec T$ from the states $\vec S$ and $\vec T$, with small numbers $\epsilon_i$ that measure the mixing contribution from the i-th family and show a hierarchy $\epsilon_3 \ll \epsilon_2 \ll \epsilon_1 \ll 1$ as needed to understand the observed hierarchy in the CKM matrix\cite{bodopmns}.  


Finally, the question: are there limitations of the approach? Certainly yes. The calculations presented are leading order with respect to many types of corrections:\\ 
-For one, a linear approximation has been used throughout for the fluctuations $\delta$ describing the quark and lepton states. Probably there will be important higher order corrections, for example effects from the heavy quarks on the light families. More concretely, next-to-leading effects from the large DM-couplings may overwhelm the tiny contributions from torsional interactions on the first family.\\
-Secondly, the calculations in this work include only {\it intra}-tetrahedral interactions of tetron isospins, i.e. interactions within one tetrahedron. Although an attempt has been made in connection with (\ref{all37}), {\it inter}-tetrahedral interactions may not have been fully taken into account. In other words, besides the top quark contributions (\ref{all37}) there may be other inter-tetrahedral effects from the lighter fermions.

It is an interesting question whether there are phenomenological predictions which distinguish the present model from other BSM theories and from the SM. First of all, it must be noted that since the model has been constructed to have a low energy limit which is as close to the SM as possible, many of its BSM effects are suppressed by negative powers of the Planck mass. However, I can see 2 major areas which go beyond the standard predictions:\\
-the parameter-free formula (98) for the PMNS matrix. In contrast to the CKM result (94) which strongly depends on running quark mass values chosen, its theoretical error is expected to be very small. As future neutrino experiments become more and more accurate, discrepancies to (98) would shake the validity of the microscopic model.\\
-the other area is cosmology. In previous work it was pointed out that the accelerated expansion of the universe obtained in the present model differs from the cosmological standard model (CSM), at least in very large time scales, and that an improved observation of the dark energy effect may be able to discriminate between the 2 models. These considerations were published before the so called `Hubble tension' (saying that measurements at different time scales provide contradictory values for the cosmological constant) was found, and it will be interesting to see whether the present model can contribute to its solution.

\vspace{3mm}

\begin{center}
{\bf Appendix A: Mathematica Program to calculate the Quark and Lepton Masses and Eigenstates}
\end{center}

The following code allows to calculate quark and lepton masses and eigenstates, given the isospin couplings as defined in the main text. The resulting masses can be found at the bottom line of the program (in GeV). 

The program's output for the eigenstates is not included in the code, but presented in (\ref{allnn5}), (\ref{allee5}), (\ref{allup}) and (\ref{alldown}). 

\noindent\(
\pmb{}\\
\pmb{\text{s10}\text{:=}\{-1,-1,-1\}\left/\sqrt{3}\right.}\\
\pmb{\text{del1u}\text{:=}\{\text{d1x},\text{d1y},\text{d1z}\}*\text{ef}}\\
\pmb{\text{del2u}\text{:=}\{\text{d2x},-\text{d2y},-\text{d2z}\}*\text{ef}}\\
\pmb{\text{del3u}\text{:=}\{-\text{d3x},\text{d3y},-\text{d3z}\}*\text{ef}}\\
\pmb{\text{del4u}\text{:=}\{-\text{d4x},-\text{d4y},\text{d4z}\}*\text{ef}}\\
\pmb{}\\
\pmb{\text{t10}\text{:=}+\text{s10}}\\
\pmb{\text{eel1u}\text{:=}\{\text{e1x},\text{e1y},\text{e1z}\}*\text{ef}}\\
\pmb{\text{eel2u}\text{:=}\{\text{e2x},-\text{e2y},-\text{e2z}\}*\text{ef}}\\
\pmb{\text{eel3u}\text{:=}\{-\text{e3x},\text{e3y},-\text{e3z}\}*\text{ef}}\\
\pmb{\text{eel4u}\text{:=}\{-\text{e4x},-\text{e4y},\text{e4z}\}*\text{ef}}\\
\pmb{}\\
\pmb{\text{dd1}\text{:=}\text{del2u}+\text{del3u}+\text{del4u}-3*\text{del1u}}\\
\pmb{\text{dd2}\text{:=}\text{del1u}+\text{del3u}+\text{del4u}-3*\text{del2u}}\\
\pmb{\text{dd3}\text{:=}\text{del1u}+\text{del2u}+\text{del4u}-3*\text{del3u}}\\
\pmb{\text{dd4}\text{:=}\text{del1u}+\text{del2u}+\text{del3u}-3*\text{del4u}}\\
\pmb{}\\
\pmb{\text{ed1}\text{:=}\text{eel2u}+\text{eel3u}+\text{eel4u}-3*\text{del1u}}\\
\pmb{\text{ed2}\text{:=}\text{eel1u}+\text{eel3u}+\text{eel4u}-3*\text{del2u}}\\
\pmb{\text{ed3}\text{:=}\text{eel1u}+\text{eel2u}+\text{eel4u}-3*\text{del3u}}\\
\pmb{\text{ed4}\text{:=}\text{eel1u}+\text{eel2u}+\text{eel3u}-3*\text{del4u}}\\
\pmb{}\\
\pmb{\text{de1}\text{:=}\text{del2u}+\text{del3u}+\text{del4u}-3*\text{eel1u}}\\
\pmb{\text{de2}\text{:=}\text{del1u}+\text{del3u}+\text{del4u}-3*\text{eel2u}}\\
\pmb{\text{de3}\text{:=}\text{del1u}+\text{del2u}+\text{del4u}-3*\text{eel3u}}\\
\pmb{\text{de4}\text{:=}\text{del1u}+\text{del2u}+\text{del3u}-3*\text{eel4u}}\\
\pmb{}\\
\pmb{\text{ee1}\text{:=}\text{eel2u}+\text{eel3u}+\text{eel4u}-3*\text{eel1u}}\\
\pmb{\text{ee2}\text{:=}\text{eel1u}+\text{eel3u}+\text{eel4u}-3*\text{eel2u}}\\
\pmb{\text{ee3}\text{:=}\text{eel1u}+\text{eel2u}+\text{eel4u}-3*\text{eel3u}}\\
\pmb{\text{ee4}\text{:=}\text{eel1u}+\text{eel2u}+\text{eel3u}-3*\text{eel4u}}\\
\pmb{}\\
\pmb{\text{vdd1}\text{:=}-2*\text{dd1}+2*\text{dd1}.\text{s10}*\text{s10}}\\
\pmb{\text{vdd2}\text{:=}-2*\text{dd2}+2*\text{dd2}.\text{s10}*\text{s10}}\\
\pmb{\text{vdd3}\text{:=}-2*\text{dd3}+2*\text{dd3}.\text{s10}*\text{s10}}\\
\pmb{\text{vdd4}\text{:=}-2*\text{dd4}+2*\text{dd4}.\text{s10}*\text{s10}}\\
\pmb{}\\
\pmb{\text{ved1}\text{:=}-2*\text{ed1}+2*\text{ed1}.\text{s10}*\text{s10}}\\
\pmb{\text{ved2}\text{:=}-2*\text{ed2}+2*\text{ed2}.\text{s10}*\text{s10}}\\
\pmb{\text{ved3}\text{:=}-2*\text{ed3}+2*\text{ed3}.\text{s10}*\text{s10}}\\
\pmb{\text{ved4}\text{:=}-2*\text{ed4}+2*\text{ed4}.\text{s10}*\text{s10}}\\
\pmb{}\\
\pmb{\text{vde1}\text{:=}-2*\text{de1}+2*\text{de1}.\text{s10}*\text{s10}}\\
\pmb{\text{vde2}\text{:=}-2*\text{de2}+2*\text{de2}.\text{s10}*\text{s10}}\\
\pmb{\text{vde3}\text{:=}-2*\text{de3}+2*\text{de3}.\text{s10}*\text{s10}}\\
\pmb{\text{vde4}\text{:=}-2*\text{de4}+2*\text{de4}.\text{s10}*\text{s10}}\\
\pmb{}\\
\pmb{\text{vee1}\text{:=}-2*\text{ee1}+2*\text{ee1}.\text{s10}*\text{s10}}\\
\pmb{\text{vee2}\text{:=}-2*\text{ee2}+2*\text{ee2}.\text{s10}*\text{s10}}\\
\pmb{\text{vee3}\text{:=}-2*\text{ee3}+2*\text{ee3}.\text{s10}*\text{s10}}\\
\pmb{\text{vee4}\text{:=}-2*\text{ee4}+2*\text{ee4}.\text{s10}*\text{s10}}\\
\pmb{}\\
\pmb{\text{ss}\text{:=}-10.70000000000000000}\\
\pmb{\text{st}\text{:=}-0.07700000000000000}\\
\pmb{\text{tt}\text{:=}-0.22000000000000000}\\
\pmb{\text{jss}\text{:=}0.32000000000000000}\\
\pmb{\text{jtt}\text{:=}0.01020000000000000}\\
\pmb{\text{jst}\text{:=}0.01750000000000000}\\
\pmb{\text{ff}\text{:=}0.00049000000000000}\\
\pmb{\text{gg}\text{:=}0.00113000000000000}\\
\pmb{\text{fg}\text{:=}0.00008500000000000}\\
\pmb{\text{ne}\text{:=}-0.00000000000103000}\\
\pmb{\text{nm}\text{:=}-0.00000000000790000}\\
\pmb{\text{nt}\text{:=}0.00000000001350000}\\
\pmb{}\\
\pmb{\text{ndd1}\text{:=}-2*\text{del1u}+2*\text{del1u}.\text{s10}*\text{s10}}\\
\pmb{\text{ndd2}\text{:=}-2*\text{del2u}+2*\text{del2u}.\text{s10}*\text{s10}}\\
\pmb{\text{ndd3}\text{:=}-2*\text{del3u}+2*\text{del3u}.\text{s10}*\text{s10}}\\
\pmb{\text{ndd4}\text{:=}-2*\text{del4u}+2*\text{del4u}.\text{s10}*\text{s10}}\\
\pmb{}\\
\pmb{\text{nee1}\text{:=}-2*\text{eel1u}+2*\text{eel1u}.\text{s10}*\text{s10}}\\
\pmb{\text{nee2}\text{:=}-2*\text{eel2u}+2*\text{eel2u}.\text{s10}*\text{s10}}\\
\pmb{\text{nee3}\text{:=}-2*\text{eel3u}+2*\text{eel3u}.\text{s10}*\text{s10}}\\
\pmb{\text{nee4}\text{:=}-2*\text{eel4u}+2*\text{eel4u}.\text{s10}*\text{s10}}\\
\pmb{}\\
\pmb{\text{zx1}\text{:=} }\\
\pmb{\text{Coefficient}[\text{ss}*(2*\text{Cross}[\text{s10},\text{dd1}]+ i *\text{vdd1})+}\\
\pmb{\text{nt}*(2*\text{Cross}[\text{s10},\text{del1u}]+ i *\text{ndd1})}\\
\pmb{+\text{st}*(2*\text{Cross}[\text{s10},\text{ed1}]+ i *\text{ved1})}\\
\pmb{+\text{jss}*\text{Cross}[\text{s10},\text{dd1}]+\text{jst}*\text{Cross}[\text{s10},\text{ed1}]+\text{nm}*\text{Cross}[\text{s10},\text{del1u}]}\\
\pmb{+ i *\text{ff}*\text{dd1}+ i *\text{fg}*\text{ed1}+ i *\text{ne}*\text{del1u},\text{ef},1]}\\
\pmb{\text{zx2}\text{:=}}\\
\pmb{\text{Coefficient}[\text{ss}*(2*\text{Cross}[\text{s10},\text{dd2}]+ i *\text{vdd2})+}\\
\pmb{\text{nt}*(2*\text{Cross}[\text{s10},\text{del2u}]+ i *\text{ndd2})}\\
\pmb{+\text{st}*(2*\text{Cross}[\text{s10},\text{ed2}]+ i *\text{ved2})}\\
\pmb{+\text{jss}*\text{Cross}[\text{s10},\text{dd2}]+\text{jst}*\text{Cross}[\text{s10},\text{ed2}]+\text{nm}*\text{Cross}[\text{s10},\text{del2u}]}\\
\pmb{+ i *\text{ff}*\text{dd2}+ i *\text{fg}*\text{ed2}+ i *\text{ne}*\text{del2u},\text{ef},1]}\\
\pmb{\text{zx3}\text{:=}}\\
\pmb{\text{Coefficient}[\text{ss}*(2*\text{Cross}[\text{s10},\text{dd3}]+ i *\text{vdd3})+}\\
\pmb{\text{nt}*(2*\text{Cross}[\text{s10},\text{del3u}]+ i *\text{ndd3})}\\
\pmb{+\text{st}*(2*\text{Cross}[\text{s10},\text{ed3}]+ i *\text{ved3})}\\
\pmb{+\text{jss}*\text{Cross}[\text{s10},\text{dd3}]+\text{jst}*\text{Cross}[\text{s10},\text{ed3}]+\text{nm}*\text{Cross}[\text{s10},\text{del3u}]}\\
\pmb{+ i *\text{ff}*\text{dd3}+ i *\text{fg}*\text{ed3}+ i *\text{ne}*\text{del3u},\text{ef},1]}\\
\pmb{\text{zx4}\text{:=}}\\
\pmb{\text{Coefficient}[\text{ss}*(2*\text{Cross}[\text{s10},\text{dd4}]+ i *\text{vdd4})+}\\
\pmb{\text{nt}*(2*\text{Cross}[\text{s10},\text{del4u}]+ i *\text{ndd4})}\\
\pmb{+\text{st}*(2*\text{Cross}[\text{s10},\text{ed4}]+ i *\text{ved4})}\\
\pmb{+\text{jss}*\text{Cross}[\text{s10},\text{dd4}]+\text{jst}*\text{Cross}[\text{s10},\text{ed4}]+\text{nm}*\text{Cross}[\text{s10},\text{del4u}]}\\
\pmb{+ i *\text{ff}*\text{dd4}+ i *\text{fg}*\text{ed4}+ i *\text{ne}*\text{del4u},\text{ef},1]}\\
\pmb{}\\
\pmb{\text{zx5}\text{:=}\text{Coefficient}[\text{st}*(2*\text{Cross}[\text{s10},\text{de1}]+ i *\text{vde1})}\\
\pmb{+\text{tt}*(2*\text{Cross}[\text{s10},\text{ee1}]+ i *\text{vee1})+\text{nt}*(2*\text{Cross}[\text{s10},\text{eel1u}]+ i *\text{nee1})}\\
\pmb{+\text{jst}*\text{Cross}[\text{s10},\text{de1}]+\text{jtt}*\text{Cross}[\text{s10},\text{ee1}]+\text{nm}*\text{Cross}[\text{s10},\text{eel1u}]}\\
\pmb{+ i *\text{gg}*\text{ee1}+ i *\text{fg}*\text{de1}+ i *\text{ne}*\text{eel1u},\text{ef},1]}\\
\pmb{\text{zx6}\text{:=}\text{Coefficient}[\text{st}*(2*\text{Cross}[\text{s10},\text{de2}]+ i *\text{vde2})}\\
\pmb{+\text{tt}*(2*\text{Cross}[\text{s10},\text{ee2}]+ i *\text{vee2})+\text{nt}*(2*\text{Cross}[\text{s10},\text{eel2u}]+ i *\text{nee2})}\\
\pmb{+\text{jst}*\text{Cross}[\text{s10},\text{de2}]+\text{jtt}*\text{Cross}[\text{s10},\text{ee2}]+\text{nm}*\text{Cross}[\text{s10},\text{eel2u}]}\\
\pmb{+ i *\text{gg}*\text{ee2}+ i *\text{fg}*\text{de2}+ i *\text{ne}*\text{eel2u},\text{ef},1]}\\
\pmb{\text{zx7}\text{:=}\text{Coefficient}[\text{st}*(2*\text{Cross}[\text{s10},\text{de3}]+ i *\text{vde3})}\\
\pmb{+\text{tt}*(2*\text{Cross}[\text{s10},\text{ee3}]+ i *\text{vee3})+\text{nt}*(2*\text{Cross}[\text{s10},\text{eel3u}]+ i *\text{nee3})}\\
\pmb{+\text{jst}*\text{Cross}[\text{s10},\text{de3}]+\text{jtt}*\text{Cross}[\text{s10},\text{ee3}]+\text{nm}*\text{Cross}[\text{s10},\text{eel3u}]}\\
\pmb{+ i *\text{gg}*\text{ee3}+ i *\text{fg}*\text{de3}+ i *\text{ne}*\text{eel3u},\text{ef},1]}\\
\pmb{\text{zx8}\text{:=}\text{Coefficient}[\text{st}*(2*\text{Cross}[\text{s10},\text{de4}]+ i *\text{vde4})}\\
\pmb{+\text{tt}*(2*\text{Cross}[\text{s10},\text{ee4}]+ i *\text{vee4})+\text{nt}*(2*\text{Cross}[\text{s10},\text{eel4u}]+ i *\text{nee4})}\\
\pmb{+\text{jst}*\text{Cross}[\text{s10},\text{de4}]+\text{jtt}*\text{Cross}[\text{s10},\text{ee4}]+\text{nm}*\text{Cross}[\text{s10},\text{eel4u}]}\\
\pmb{+ i *\text{gg}*\text{ee4}+ i *\text{fg}*\text{de4}+ i *\text{ne}*\text{eel4u},\text{ef},1]}\\
\pmb{}\\
\pmb{\text{S535}\text{:=}\text{Flatten}[i\{\text{zx1},\text{zx2},\text{zx3},\text{zx4},\text{zx5},\text{zx6},\text{zx7},\text{zx8}\}]}\\
\pmb{}\\
\pmb{\text{Eigenvalues}[}\\
\pmb{\{}\\
\pmb{\text{Coefficient}[\text{S535},\text{d1x},1],}\\
\pmb{\text{Coefficient}[\text{S535},\text{d1y},1],}\\
\pmb{\text{Coefficient}[\text{S535},\text{d1z},1],}\\
\pmb{\text{Coefficient}[\text{S535},\text{d2x},1],}\\
\pmb{-\text{Coefficient}[\text{S535},\text{d2y},1],}\\
\pmb{-\text{Coefficient}[\text{S535},\text{d2z},1],}\\
\pmb{-\text{Coefficient}[\text{S535},\text{d3x},1],}\\
\pmb{\text{Coefficient}[\text{S535},\text{d3y},1],}\\
\pmb{-\text{Coefficient}[\text{S535},\text{d3z},1],}\\
\pmb{-\text{Coefficient}[\text{S535},\text{d4x},1],}\\
\pmb{-\text{Coefficient}[\text{S535},\text{d4y},1],}\\
\pmb{\text{Coefficient}[\text{S535},\text{d4z},1],}\\
\pmb{\text{Coefficient}[\text{S535},\text{e1x},1],}\\
\pmb{\text{Coefficient}[\text{S535},\text{e1y},1],}\\
\pmb{\text{Coefficient}[\text{S535},\text{e1z},1],}\\
\pmb{\text{Coefficient}[\text{S535},\text{e2x},1],}\\
\pmb{-\text{Coefficient}[\text{S535},\text{e2y},1],}\\
\pmb{-\text{Coefficient}[\text{S535},\text{e2z},1],}\\
\pmb{-\text{Coefficient}[\text{S535},\text{e3x},1],}\\
\pmb{\text{Coefficient}[\text{S535},\text{e3y},1],}\\
\pmb{-\text{Coefficient}[\text{S535},\text{e3z},1],}\\
\pmb{-\text{Coefficient}[\text{S535},\text{e4x},1],}\\
\pmb{-\text{Coefficient}[\text{S535},\text{e4y},1],}\\
\pmb{\text{Coefficient}[\text{S535},\text{e4z},1]}\\
\pmb{\}}\\
\pmb{]}\)

\noindent\(\pmb{\{170.794,170.794,170.794,4.35497,4.35497,4.35497,1.74351,}\\
\pmb{1.33497,1.33497,1.33497,0.10551,0.097825,0.097825,0.097825,}\\
\pmb{0.00477782,0.00477782,0.00477782,0.00221218,0.00221218,}\\
\pmb{0.00221218,0.00051,4.7123*10{}^{\wedge}-11,8.92766*10{}^{\wedge}-12,1.02624*10{}^{\wedge}-12\}}\)

\vspace{3mm}

\begin{center}
{\bf Appendix B: How to include the Strong Interaction in the Tetron Scheme}
\end{center}

As shown in a series of papers\cite{bodoreview,bodogravity,bodomasses,bodotalk,bodohiggs} the properties of gravity and of electroweak processes can be reduced to properties of tetrons. The present calculations are an example for this statement proving explicitly that the fermion mass spectrum and family mixings can be correctly obtained within the microscopic model. 

In these calculations each quark flavor appears as a triplet of the Shubnikov group $G_4$, eq. (\ref{eq8gs}), with 3 degenerate masses. But how does the strong interaction, where quarks usually are considered as SU(3) triplets, fit into the tetron picture?

The dominant features of the strong interaction are the linear attractive potential at low energies and asymptotic freedom at high energies. In the tetron model the strong interaction is related to disturbances by the triplet isospin excitations(=quarks) of the local ground state which is formed by a single tetrahedron of isospin vectors. As triplet states of $G_4$, quarks disturb the ground state's isomagnetism, whereas leptons are $G_4$-singlets, i.e. `isomagnetically' neutral. They do not disturb the ground state and can exist freely, not taking part in the strong interaction. 

As discussed in \cite{bodoreview}, the isomagnetic ground state energy of 2 neighboring tetrahedrons is roughly $E_{QCD}\approx 1$ GeV corresponding to a characteristic length scale $L_{QCD}\approx 10^{-15}$m. The linear potential between two $G_4$-triplets, vulgo a quark $Q$ and an antiquark $\bar Q$, then arises as follows: Since the inter-tetrahedral exchange energy $j=E_{QCD}$ is relatively small, its physical effects have a much longer range $L_{QCD}$ than the weak interactions which are induced by the inner-tetrahedral exchange energies $J=O(100)$ GeV. The triplet excitations corresponding to the two quarks are characterized by small vibrations $\vec \delta_Q$ and $\vec \delta_{\bar Q}$ of the isospin vectors (\ref{eq894}). When the distance between the two excitations becomes larger than $L_{QCD}$, an additional pair $\vec \delta_q$ and $\vec \delta_{\bar q}$ is excited on intermediate tetrahedrons in order to reduce the original `isomagnetic' suspense between $\vec \delta_Q$ and $\vec \delta_{\bar Q}$.  The associated cost in energy is proportional to the number of $q\bar q$ pairs created, and the potential V between $Q$ and $\bar Q$ therefore increases linearly with distance:
\begin{equation}
V=F |x|  \qquad \qquad \qquad  F\approx -j \langle \vec\delta_Q \vec\delta_{\bar Q} \rangle /L_Q
\label{aba:eqvv}
\end{equation}
where $\langle \vec\delta_Q \vec\delta_{\bar Q} \rangle$ is the isospin correlation between the sites, on which the isospins vibrate, and $L_{QCD}$ the length where all this becomes relevant. The confinement energy is hence proportional to the original `ferromagnetic' exchange energy j induced on the disturbances $Q$ and $\bar Q$. The ratio $x/L_{QCD}$ is the number of times, an additional pair of excitations has to be created from the 'sea'.

In the tetron model quarks are disturbances $\vec \delta_Q$ of the isospin vectors (\ref{eq894}). Due to isospin interactions like (\ref{mm3}) not only the ground state vectors but also the disturbances tend to align. This tendency of the triplet excitations gives rise to a `mass gap' $\langle \vec\delta_Q \vec\delta_{\bar Q} \rangle \neq 0$ which signals a phase transition in the form of the usual breakdown of chiral symmetry due to the strong interactions.

In summary, a single quark $Q$ increases the energy of the system in its neighborhood $L_{QCD}$ not only by its flavor-dependent mass(=excitation energy) but by an additional energy necessary to `pick up a $q\bar q$ pair from the sea'. This energy is flavor independent, because it does not depend on the flavor Q, which flavors q are excited. The flavors q correspond to an average of the light quarks u, d and s. So when a $Q$ and a $\bar Q$ are torn apart, at some distance $x\approx L_{QCD}$ a light $q \bar q$ pair is formed, because otherwise the single quark $Q$ could not endure the disturbance of the ground state. In the end a sort of string appears obtained by $Q \bar q q \bar q' ... \bar Q$ pairs. Any time a new $q\bar q$ pair is created, energy is to be taken from the environment, so the associated cost in energy is proportional to the number of $q\bar q$ pairs and the potential between quark and antiquark therefore increases linearly with distance as indicated in (\ref{aba:eqvv}).


Readers familiar with the strong interaction, will recognize that one is led this way to the classic ideas of the quark model. For example, using the linear potential (\ref{aba:eqvv}) masses of mesons and baryons can be estimated just as in the quark model. Since mesons and baryons are $G_4$-singlets, the `isomagnetic' disturbances induced by quarks get neutralized in these bound states, i.e. mesons and baryons do not disturb the ground state of a single tetrahedron. 

The role of the length $L_{QCD}$, where the creation of a light quark-antiquark is enforced, is the same as in the Standard Model. At distances above $L_{QCD}$ one has confinement, while below $L_{QCD}$ the strong force diminishes. Virtual bound quark-antiquark pairs are formed which as gluons mediate a strong interaction of the original $Q\bar Q$ pair which effectively can be described by the QCD Lagrangian with its local SU(3) gauge symmetry. As well known, this interaction dies out when the energies involved go to infinity, i.e. one has asymptotic freedom.

\vspace{3mm}

\begin{center}
{\bf Appendix C: The SM Boson Sector}
\end{center}

\begin{center}
\begin{table}
\begin{center}
\begin{tabular}{ | l | l |}
\hline
Photon $\gamma$ & Higgs Particle H \\
\hline
$\delta \, \mathfrak{Re} (U_1 + U_2)$ & $\delta \, \mathfrak{Re} (U_1-U_2)$ \\
\hline
$S=1$ & $S=0$ \\
$I=0$ & $I=0$ \\
$P=0$ (in phase, massless) & $P=1$ \\
\hline\hline
$\vec W$ Bosons & $\vec \xi$ Goldstone modes \\
\hline
$\delta \, \mathfrak{Re} (D_1- D_2)$& $\delta \, \mathfrak{Re} (D_1+ D_2)$\\
$\delta \, \mathfrak{Im} (D_1- D_2)$& $\delta \, \mathfrak{Im} (D_1+ D_2)$\\
$\delta \, \mathfrak{Im} (U_1- U_2)$& $\delta \, \mathfrak{Im} (U_1+ U_2)$ \\
\hline
$S=1$ & $S=0$ \\
$I=1$ & $I=1$ \\
$P=1$ & $P=0$ (in phase, massless) \\
\hline
\end{tabular}
\caption{Bosonic excitation states of the two tetrons (\ref{pm11h1}) as allowed by the Pauli principle, and their identification with the boson content of the Standard Model. S and I denote spin and isospin, respectively, and P an additional quantum number which describes (anti)symmetry of the wave function under the exchange of the two tetrons involved. P is defined after (\ref{pm11h1}) and used to establish the Pauli principle for the 2-tetron excitations. The octonion representation (\ref{eq8}) describing a tetron field $\Psi$ comprises particle and antiparticle degrees of freedom within one representation, so that actually one may talk about excitations of a pair of 2 tetrons - the reason why the Pauli principle is applicable in order to classify those excitations.\\
Not counting spin dof, there are 8 isospin degrees of freedom to describe the SM bosons, as provided by (\ref{pm114}). Refraining from spin is allowed since we consider excitations in isospin space only, while the tetrons remain unexcited with respect to spin in physical space. $\vec \xi$ and $\gamma$ are massless and can be interpreted as Goldstone modes, as explained in the text.}
\end{center}
\end{table}
\end{center}



In the main part of this work excitations of tetrons {\it within one} of the internal tetrahedrons are considered and the quark lepton spectrum (6 singlets and 6 triplets) is obtained from the tetrahedral Shubnikov symmetry group $G_4$. Mass formulas are given in terms of isospin couplings among tetrons and unambiguous predictions for the CKM and PMNS mixing matrices are obtained. The spin-$\frac{1}{2}$ nature of the excitations is guaranteed by the Dirac nature of tetrons (\ref{eq8}) in the base space (i.e. in physical space).

In contrast, the SM Higgs and gauge boson fields are to be considered as excitations of two tetrons $\Psi_1$ and $\Psi_2$ belonging to {\it two neighboring} tetradedrons 1 and 2. Their masses thus must arise from {\it inter}-tetrahedral isospin interactions (while quark and lepton masses are due to {\it inner}-tetrahedral ones), and their spin-0 and spin-1 nature from tensor products of the two spin-$\frac{1}{2}$ tetrons sitting on two different base points of physical space. As excitations of single pairs they do not feel their tetrahedral environment, which means they transform under isospin but do not appear in multiplets of $G_4$. Rather they can be constructed as vibrations $\delta$ around the ground states. According to (\ref{eq8}) a tetron forms an isospin doublet, so that one may write
\begin{eqnarray}
\Psi_1=
\begin{bmatrix}
\delta D_1 \\
\langle U \rangle + \delta U_1 \\
\end{bmatrix}
\qquad \qquad
\Psi_2=
\begin{bmatrix}
\delta D_2 \\
\langle U \rangle + \delta U_2 \\
\end{bmatrix}
\label{pm112}
\end{eqnarray}
where the ground states $\langle\Psi_{1,2} \rangle=(0,\langle U \rangle)$ are aligned radial isospins pointing outward along the internal z-direction and are assumed to combine to a spin singlet 2-tetron ground state.

In recent papers it was claimed that the bosonic sector of the microscopic model naturally comes out as a 2-Higgs Doublet Model (2HDM)\cite{branco,bodohiggs}. However, in the following it will be shown that one can as well construct a version of the microscopic model so that the boson content of the SM arises, not more and not less, as given in Table 1. The point is that (\ref{pm112}) contains 8 real vibrational degrees of freedom in isospin space, and it will be seen below how these can be mapped to the 8 boson fields contained in Table 1.

In fact, if one includes interactions among the 2 excited tetrons (\ref{pm112}) one typically arrives at 8 vibrational eigenstates of the form
\begin{eqnarray}
\delta \, \mathfrak{Re} (D_1\pm D_2), \quad
\delta \, \mathfrak{Im} (D_1\pm D_2), \quad
\delta \, \mathfrak{Re} (U_1\pm U_2), \quad
\delta\, \mathfrak{Im} (U_1\pm U_2)
\label{pm114}
\end{eqnarray}
The ones with the plus sign turn out to be massless. As explained below they can be interpreted as Goldstone modes.



\begin{figure}[h]
\begin{center}
\includegraphics[width=3.1in]{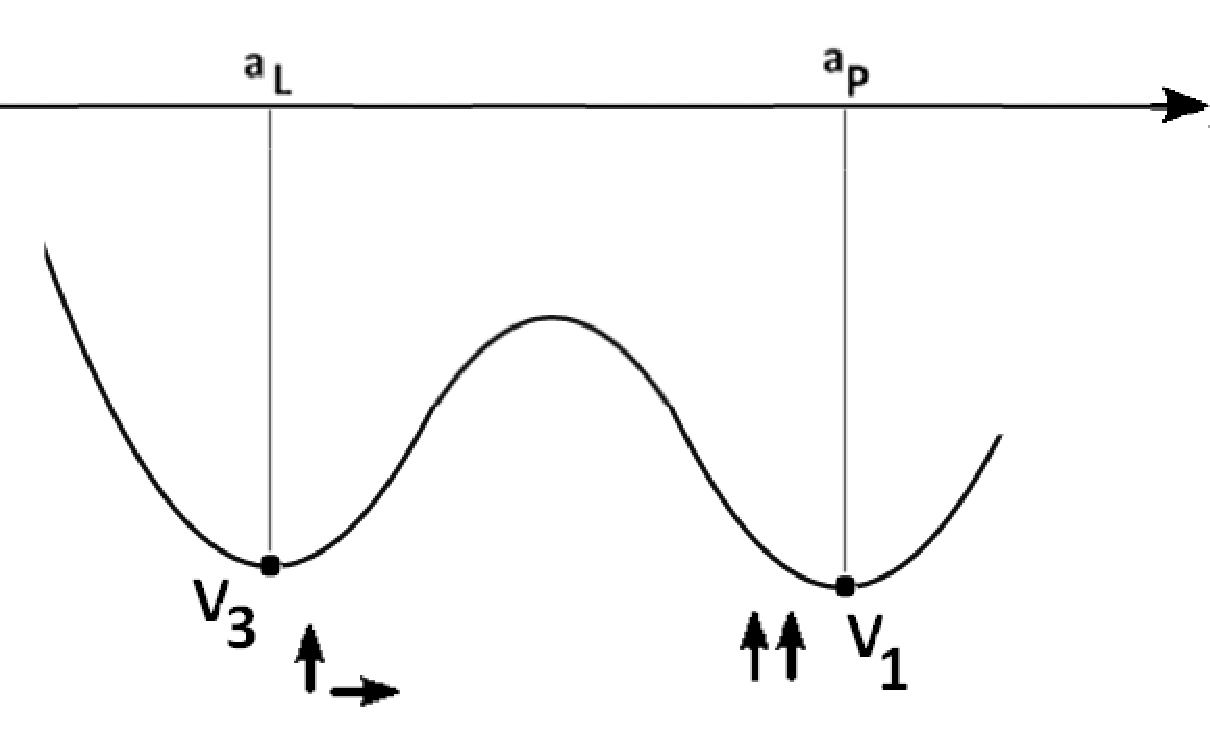}
\end{center}
\caption{The total isomagnetic potential V between 2 tetrons as a function of their distance. The Planck length $a_P$ is defined in Fig. 1 as the distance between 2 neighboring tetrahedrons, while $a_L$ corresponds to the extension of any one tetrahedron in the 3 extra dimensions. The minimum of V at $a_P$ essentially is due to the Heisenberg singlet potential $V_1$. $V_1$ is the dominant contribution to the {\it inter}-tetrahedral interactions and causes neighboring tetrahedrons of isospin vectors to align as shown in Fig. 1. On the other hand, the minimum at $a_L$ arises from the DM triplet potential $V_3$ introduced in (\ref{eqrt15d}). It dominates the {\it inner}-tetrahedral interactions and gives the top quark its mass in the sense that $m_t \sim K_{LL} \sim V_3(a_L)$, cf. (\ref{all36}).
The values $V_3(a_L)$ and $V_1(a_P)$ are naturally of the same order, because they both stem from one original potential $W_7$ defined in connection with (\ref{eq19}) and (\ref{eqrt15a}). Note that $ J_{inner} :=V_1(a_L)\sim 1$ GeV is much smaller than $ J_{inter}:= V_1(a_P)\sim 100$ GeV and according to (\ref{all36}) is responsible for the mass of the second family.}
\end{figure}

The calculation of quark and lepton masses in Sect. IV proves the dominance of the DM interactions within each tetrahedron, so that actually each internal tetrahedron is a (frustrated) DM isomagnet, cf. the discussion after (\ref{all36}). In contrast, DM must not play a role in the `inter-tetrahedral' interactions of tetrons from two different tetrahedrons, because DM prefers isospins at 90 degrees while all tetrahedrons are aligned in parallel after the SSB according to Fig. 1, and this presupposes dominance of the aligning inter-tetrahedral Heisenberg interactions.

In Sect. III singlet and triplet potentials $V_1$ and $V_3$ have been identified as the source of the Heisenberg and DM tetron interactions, respectively. As shown in Fig. 3, the inter-tetrahedral dominance of $V_1$ corresponds to an energy minimum at Planck length $a_P$, while the inner-tetrahedral dominance of $V_3$ corresponds to an energy minimum at $a_L$. Both minima naturally have the same order of magnitude $V_{1min} \sim V_{3min}$ because both $V_1$ and $V_3$ stem from one and the same underlying potential $W_7$ according to (\ref{eqrt15a}). This explains why $m_t$ [dominated by $K_{LL}=V_3(a_L)$in (\ref{all36})] and $m_H$ [dominated by $V_1(a_P)$] roughly have the same order of magnitude.

Assume for a moment that instead of excitations one would have bound states of 2 tetrons (\ref{pm112}) sitting in 2 neigboring tetrahedrons, and furthermore a non-relativistic situation. Then with the notation used in (\ref{eqrt198}) these bound states would transform as
\begin{eqnarray}
(2,2)\times (2,2)=(1+3,1+3)=(1,1)+(1,3)+(3,1)+(3,3)
\label{pm114ds}
\end{eqnarray}
and one would be tempted to interpret (1,1+3) as the Higgs doublet and (3,1) and (3,3) as a massive non-relativistic `photon' and a $\vec W$ field. However, instead of bound states one has 2-tetron vibrational excitations within the 3 isospin extra dimensions. Therefore, forming a tensor product is only reasonable in the base physical space, where there are no vibrations and the states transform as $2\times 2=1+3$ under spin rotations, i.e. they are bosons with either spin 0 or spin 1 and accordingly are listed in Table 1.

Concerning the vibrations in isospin space one must not consider tensor products but is obviously led to the representation (\ref{pm112}) so that the SM Higgs in terms of tetron vibrations reads
\begin{eqnarray}
\langle U \rangle
\begin{bmatrix}
\delta (D_1+D_2) \\
-2\mathfrak{Re} \delta (U_1-U_2)-2\mathfrak{Im} \delta (U_1+U_2) \\
\end{bmatrix}
=v_F^2
\begin{bmatrix}
\xi_1 +i \xi_2 \\
v_F + H + i \xi_3 \\
\end{bmatrix}
\label{pm113}
\end{eqnarray}
The Heisenberg interaction between the isospin vectors
\begin{eqnarray}
Q_{1,2}=\frac{1}{2}\Psi_{1,2}^\dagger \vec \tau\Psi_{1,2}
\label{pm11gg77}
\end{eqnarray}
belonging to the tetrons $\Psi_1$ and $\Psi_2$ in (\ref{pm112}) amounts to
\begin{eqnarray}
\frac{d\vec Q_1}{dt}=J_{inter} \vec Q_2 \times \vec Q_1 \qquad \qquad
\frac{d\vec Q_2}{dt}=J_{inter} \vec Q_1 \times \vec Q_2
\label{pm11gg}
\end{eqnarray}
with the inter-tetrahedral Heisenberg coupling $J_{inter}$ determined by the singlet potential $V_1(a_P)$ as described in connection with Fig. 3. There are then massless eigenmodes $\vec Q_1+\vec Q_2$ as well as massive ones $\vec Q_1-\vec Q_2$.

Terms including $\gamma_5$ need not be considered here. As discussed in \cite{bodoreview} and before (\ref{dig55}), chiral interactions are not a fundamental property of tetrons but a consequence of the presence of the parity violating tetrahedral star configuration of isospin vectors, and inclusion of $\gamma_5$ in $\vec Q_{L,R}$ for the calculation of the quark and lepton masses is only a technical means to cover all the 24 dof of the inner-tetrahedral vibrations.

Instead of discussing it on the level of isospin vectors, the dynamics in (\ref{pm11gg}) should be rewritten in terms of the quantities in (\ref{pm114}). The point is that in order to cover all bosonic dof appearing in Table 1 not only vibrations of the isospin vectors but also of the isospin densities have to be taken into account - altogether these dof are given in (\ref{pm114}). Eq. (\ref{pm11gg}), however, does not determine the U but only the D  components according to
\begin{eqnarray}
\frac{d}{dt} \mathfrak{Re} (\delta D_1 - \delta D_2) &=& 2 J_{inter} 
\mathfrak{Im} (\delta D_1 -\delta D_2) \nonumber \\
\frac{d}{dt} \mathfrak{Im} (\delta D_1 -\delta D_2) &=& -2 J_{inter} 
\mathfrak{Re} (\delta D_1 -\delta D_2)
\label{pm11h1}
\end{eqnarray}
corresponding to 2 modes of equal mass $2 J_{inter}$, and to be interpreted as the mass of the $W^\pm$ bosons.

Furthermore, there are 2 massless modes $\mathfrak{Re} \delta (D_1 + D_2)$
and $\mathfrak{Im} \delta (D_1 + D_2)$ corresponding to the modes $\xi_1$ and $\xi_2$, cf. Table 1 and (\ref{pm113}). They are reminiscent to the appearance of massive so called `optical' and massless so called `acoustic' modes in lattice vibrations where the acoustic phonons can be interpreted as Goldstone particles corresponding to the broken translational symmetry. Microscopically, these Goldstone modes correspond to in-phase vibrations of 2 atoms in the primitive cell of the crystal, while the massive `optical' modes are anti-phase, i.e. with an antisymmetric wave fuction.

Similarly, in the present case, the vibrations introduced in (\ref{pm114}) with a plus sign are massless and in-phase, corresponding to a symmetric wave function, whereas those with a minus sign have an antisymmetric wave function. The wave functions being symmetric or anti-symmetric, one can then analyze the possible vibrational states with the help of the Pauli principle applied to 2 tetrons and obtains the allowed states in Table 1.

Just as phonons from 2 atoms in the primitive cell, the bosonic excitations of 2 tetrons in neighboring tetrahedrons move as quasiparticle waves on physical space, and the massless $\vec\xi$-fields serve as Goldstone modes of the broken local isospin symmetry. 

Finally it should be stressed that results only for the $\delta D$ components of the tetrons (\ref{pm112}) have been obtained here.
For vibrations of the $\delta U$ components one should go back to the definition of the exchange integrals J and find that their coefficient is not just the Heisenberg product $\vec Q_1 \vec Q_2$ but in addition includes a term involving the tetron densities, so that the exchange energy is generically of the form $J_V \vec Q_1 \vec Q_2 + J_S n_1 n_2$\cite{auerbach}, where $J_V:=V_{1V}(a_p)$ is the exchange integral for the vector and $J_S:=V_{1S}(a_P)$ for the scalar. This then gives the U-excitations corresponding to the SM Higgs, $W_3$ and $B$ field. More precisely, excitations of $Q_1^x Q_2^x$ and $Q_1^y Q_2^y$ lead to the $W^\pm$ mass, excitations of $Q_1^z Q_2^z$ to the Z mass and of $n_1 n_2$ to the Higgs mass. 

One obtains the Weinberg mixing of the SM $W_3$ and $B$ boson fields by including a complex vev for the U-tetron, i.e. $\langle U \rangle\sim \exp {i\theta/4}$ where $\theta=\arccos(-\frac{1}{3})$ is the tetrahedral angle of about 109 degrees and the Weinberg angle turns out to be $\theta/4$. Note the appearance of a complex phase in $\langle U \rangle$ is not only allowed (because it does not change the reality of the SM vev $\langle \bar U U\rangle$) but is also mandatory in order that the complete wave function for the tetrahedral ground state corresponds to a standing wave around the 4 tetrons in the sense of the Bohr Sommerfeld quantization rules.

Thus in their ground state tetrons $\langle U_1 \rangle$ and $\langle U_2 \rangle$ sitting in neighboring corners of neighboring tetrahedrons are in phase. However, their particle and anti-particle components differ in phase by $\theta/2$.

\vspace{3mm}

\begin{center}
{\bf Appendix D: The SO(7,1) Choice}
\end{center}

In this appendix an alternative starting point of the microscopic model is briefly discussed, where a 7+1 dimensional spacetime is considered instead of a 6+1 dimensional one. As shown below, considering SO(7,1) does not change the calculations and conclusions drawn in this work, but has some conceptual advantages over SO(6,1).

In 7+1 dimensions there are two complex spinorial octonion representations $8_L$ and $8_R$\cite{ross,slansky}, one with left handed and the other with right handed chirality. When decomposing $SO(7,1)$ into the Lorentz symmetry $SO(3,1)$ of the base Minkowski spacetime and the internal $SO(4)$, the expression (\ref{eq8}) gets replaced by\cite{slansky} 
\begin{eqnarray}
SO(7,1)&\rightarrow& SO(3,1)\times SU(2) \times SU(2) \nonumber \\
8_L &\rightarrow& ((1,2),(2_L,1))+((2,1),(1,2_R)) \nonumber \\
8_R &\rightarrow& ((1,2),(1,2_R))+((2,1),(2_L,1)) \nonumber \\
8_L \oplus 8_R &\rightarrow& ((1,2)+(2,1),(1,2_R)+(2_L,1))
\label{eq871}
\end{eqnarray}
where the covering group $SU(2)\times SU(2)$ of $SO(4)$ has been introduced. $(2,1)$ and $(1,2)$ denote left and right handed Weyl spinor representations of $SO(3,1)$ while the left and right handed $SO(4)$ spinors are denoted by $(2_L,1)$ and $(1,2_R)$. Note, the chiralities of all the doublets appearing in (\ref{eq871}) follow from the chiralities of the parent representations $8_L$ and $8_R$.

In order to obtain the necessary internal (i.e. $SO(4)$) vibrational degrees of freedom, the decompositon of $8_L \oplus 8_R$ in the last line of (\ref{eq871}) should be considered. The corresponding tetron field represents a Dirac field $(1,2)+(2,1)$ in the $SO(3,1)$ base part, with 2 $SO(4)$ doublets $\Psi_L=(U_L,D_L)$ and $\Psi_R=(U_R,D_R)$ on top. Note that here the indices L and R refer to the chiralities in the 4 extra dimensions. Accordingly, the isospin angular momenta in eq. (\ref{eq894}) should be reinterpreted with respect to the two $SU(2)$ subgroups in $SO(4)$, and not using the $\gamma_5$ of the Lorentz group but $\gamma_5^{SO(4)}$ defined in $SO(4)$:
\begin{eqnarray}
\vec Q_L=\frac{1}{4}\Psi^\dagger (1-\gamma_5^{SO(4)})\vec \tau\Psi =\frac{1}{2}\Psi^\dagger_L \vec\tau \Psi_L
\quad 
\vec Q_R= \frac{1}{4}\Psi^\dagger (1+\gamma_5^{SO(4)})\vec \tau\Psi=\frac{1}{2}\Psi^\dagger_R \vec\tau \Psi_R 
\label{eq894bb}
\end{eqnarray} 
and similarly for $n_L$ and $n_R$ in (\ref{pm11gg92}). 

On each tetrahedral site there are thus 8 independent internal vibrators, namely $\vec Q_L$, $\vec Q_R$, $n_L$ and $n_R$. A tetrahedron with 4 sites then gives 32 vibrators, 24 of which make up for the observed quarks and leptons and the remaining 8 for the dark family, as introduced after (\ref{pm11gg92}).
This point of view effectively is equivalent to having two SO(6,1) tetrons on each tetrahedral site, cf. Fig. 5 in \cite{bodoreview}.

Otherwise, much of the analysis for $SO(7,1)$ can be taken over from $SO(6,1)$, with possible complications due to the presence of more indices. For example, instead of the triplet potential (\ref{eqrt15d}) there may appear potentials with more indices transforming under $SU(2) \times SU(2)$. In addition, the fact that $G_4$ is a subgroup of only one of the 2 isospin SU(2)s may affect and simplify the derivation of weak parity violation given in \cite{bodoreview}.
Chosing $SO(7,1)$ has the further advantage that (quasi particle) waves obey Huygens' principle in 7+1 but not in 6+1 dimensions\cite{atiyah}.


\end{document}